\documentclass[journal]{IEEEtran}

\usepackage[english]{babel}
\usepackage{graphicx}
\usepackage{cite}
\usepackage{amsmath,amssymb,amsfonts}
\usepackage{mleftright}
\usepackage{booktabs} 
\usepackage[numbers,sort&compress]{natbib}
\usepackage{algorithm}
\usepackage[noend]{algpseudocode}
\usepackage{threeparttable}
\usepackage{graphicx}
\usepackage{tabularx} 
\usepackage{textcomp}
\usepackage[T1]{fontenc}
\usepackage{xcolor}
\usepackage{comment}
\usepackage{makecell}

\usepackage{caption}
\usepackage{subcaption}

\usepackage{bbold}

\ifCLASSINFOpdf
 
\else
  
\fi

\begin{document}

\title{Differentially Private Online Distributed Aggregative Games With Time-Varying and Non-Identical Communication and Feedback Delays}

\author{Olusola~Odeyomi,~\IEEEmembership{Member,~IEEE,}
~Tokunbo~Ogunfunmi,~\IEEEmembership{Senior Member, IEEE,}
and~Adjovi~Laba~\IEEEmembership{}

\thanks{O. T. Odeyomi and Adjovi Laba are with the Department
Computer Science, North Carolina Agricultural and Technical State University, Greensboro,
NC, 27411 USA e-mail: (otodeyomi@ncat.edu, anlaba@aggies.ncat.edu).}
\thanks{T. Ogunfunmi is with the Department of Electrical and Computer Engineering, Santa Clara University, Santa Clara, CA, 95053. }
}

\markboth{Journal of \LaTeX\ Class Files,~Vol.~14, No.~8, August~2015}%
{Shell \MakeLowercase{\textit{et al.}}: Bare Demo of IEEEtran.cls for IEEE Journals}

\maketitle

\begin{abstract}
This paper investigates online distributed aggregative games with time-varying cost functions, where agents are interconnected through an unbalanced communication graph. Due to the distributed and noncooperative nature of the game, some curious agents may wish to steal sensitive information from neighboring agents during parameter exchanges. Additionally, communication delays arising from network congestion, particularly in wireless settings, as well as feedback delays, can hinder the convergence of agents to a Nash equilibrium. Although a recent work addressed both communication and feedback delays in aggregative games, it is based on the unrealistic assumption that the delays are fixed over time and identical across agents. Hence, the case of time-varying and non-identical delays across agents has never been considered in aggregative games. In this work, we address the combined challenges of privacy leakage with time-varying and non-identical communication and feedback delays for the first time. We propose an online distributed dual averaging algorithm that simultaneously tackles these challenges while achieving a provably low regret bound. Our simulation result shows that the running average of each client's local action converges over time.  
\end{abstract}


\IEEEpeerreviewmaketitle

\section{Introduction}
\IEEEPARstart{A}{ggregative} games have garnered considerable interest because of their broad range of applications, including smart grid control \cite{deng2021distributed}, energy consumption \cite{xu2022efficient}, and network congestion management \cite{grammatico2017dynamic}. In this type of non-cooperative game, each agent's cost function is influenced not only by their individual action but also by some aggregate measure of all agents' actions. A common solution concept in this context is the Nash equilibrium, where no agent can benefit by changing their strategy unilaterally.

In recent years, a wide range of algorithms have been proposed for computing Nash equilibria in aggregative games, encompassing both semidecentralized approaches \cite{belgioioso2017semi,grammatico2017dynamic,de2019continuous} and fully distributed approaches \cite{parise2020distributed,koshal2016distributed,zhu2020distributed,huang2022linearly}. Semi-decentralized algorithms rely on a central coordinator to collect and disseminate aggregated information based on local player actions. However, their performance may deteriorate under network congestion or adversarial attacks targeting the coordinator. To address these limitations, fully distributed algorithms — operating without any central coordination — have attracted growing interest due to their scalability and robustness. These algorithms typically structure agents within a graph network, which may be either static \cite{koshal2012gossip,koshal2016distributed} or time-varying \cite{zuo2023distributed,belgioioso2020distributed,chen2025communication,yao2025online,fang2022distributed,zhu2023distributed,wang2022distributed}. Time-varying graphs are particularly effective for modeling wireless and mobile environments, where communication links change dynamically. Furthermore, graph networks can be categorized as balanced \cite{deng2018distributed,deng2018distributeda} or unbalanced \cite{zhang2022distributed,zuo2023distributed}. While balanced graphs assume equal in-degree and out-degree for each agent—a condition rarely met in practice — unbalanced graphs relax this constraint, making them more suitable for modeling realistic distributed systems.

Aggregative games are best modeled as an online learning problem when the learning environment is dynamic. Therefore, the online aggregative game setting allows each agent to receive sequential feedback information from the environment after selecting an action to minimize its cost function with the aim of achieving the Nash equilibrium. The dynamic and uncertain nature of the environment makes the cost function and the Nash equilibrium time-varying. An online learning algorithm designed to track the time-varying Nash equilibrium should mimic its offline counterpart. This is measured by a metric called regret. Regret can be static or dynamic. Dynamic regret is often used to track the time-varying Nash equilibrium, and a good online algorithm must achieve a sublinear regret bound. Some recent studies have been conducted on distributed online aggregative games as shown in \cite{liu2023online, liu2024sub,liu2025distributed,shokri2021network, zuo2023distributed,lin2023statistical,yao2025online}. 

Distributed systems are vulnerable to privacy breaches, particularly when a curious agent with malicious intentions steals sensitive information during communication exchange with neighbors without disrupting the normal execution of the algorithm. Common techniques used for protecting sensitive information include homomorphic encryption \cite{acar2018survey}, secure multiparty computation \cite{lindell2020secure}, and differential privacy \cite{dwork2006differential}. While homomorphic encryption and secure multiparty computation offer strong privacy guarantees, they are often computationally intensive. In contrast, differential privacy employs a simpler approach by injecting noise into data to obscure sensitive information, making it the preferred method for privacy preservation in distributed systems. Differential privacy has been extensively applied over the years in cooperative distributed settings, such as federated learning \cite{wei2020federated,truex2020ldp,odeyomi2021differentially}. 
However, its application in non-cooperative game settings remains relatively new. Some of the few works that recently applied differential privacy in non-cooperative games can be found in \cite{you2022non,wang2024differentially,chen2025differentially,chen2024communication,ye2021differentially}.

In distributed systems, communication delays can occur between any agent transmitting information to another agent due to limited bandwidth or network failures, which can degrade system performance and complicate algorithm design and analysis \cite{zhu2018limits}. Some existing work proposed distributed algorithms for non-cooperative games robust to fixed communication delays \cite{shokri2021network} or time-varying communication delays \cite{zhang2020distributed,liu2025distributed, wang2022robust}. In addition to communication delays, agents may also encounter feedback delays, where time-varying cost functions are observed only after a certain lag. As a result, the gradients used to update parameters are computed based on outdated cost functions, introducing further challenges to timely and accurate learning. Recently, Liu et al. \cite{liu2023online} investigated the effects of feedback delays in online distributed aggregative games. As a limitation, the game setting assumes a fixed feedback delay across all agents and over time. In continuation, Liu et al. \cite{liu2024sub} addressed the dual challenge of communication delays and feedback delays in online distributed aggregative games. However, the game setting assumes a fixed communication delay and a fixed feedback delay across agents and over time.  It is unrealistic to assume that all agents will encounter identical delays or that any single agent will consistently experience the same delay over time. For example, distributed wireless sensors deployed across various geographic regions are subject to differing environmental conditions, resulting in non-uniform delays across the network. Likewise, as environmental conditions change, the delay experienced by an individual agent can vary over time. Addressing time-varying and non-identical communication delays and feedback delays simultaneously has never been considered. Refer to Table 1 for a comparison of recent works.

In this paper, we address simultaneously, for the first time, the combined challenges of privacy leakage coupled with time-varying and non-identical communication delays and feedback delays in online distributed aggregative games.  We consider a time-varying unbalanced graph network. In the absence of a central coordinator, each agent iteratively computes an estimated aggregate function in a fully distributed manner.
 To tackle these challenges, we propose a novel dual averaging algorithm that ensures robustness to privacy leakage by randomizing each agent’s sensitive parameters using Laplace noise, thus preserving differential privacy. Furthermore, we incorporate time-varying and non-identical communication and feedback delays by leveraging delayed received information from neighbors and delayed gradients, which helps reduce idle times and improves the efficiency of the iterative process. To correct for the unbalanced structure of the graph, we adjust the distributed updates using an estimate of the left eigenvector corresponding to an eigenvalue of 1 for the unbalanced graph row-stochastic weight matrix. Specifically, our major contributions are as follows: 

\begin{itemize}
    \item We propose an online dual averaging algorithm that addresses the combined challenges of privacy leakage, \textit{time-varying} and \textit{non-identical} communication delays and feedback delays in online distributed aggregative games for the first time. The proposed algorithm randomizes each agent's sensitive parameters with Laplace noise before sharing them with its time-varying neighbors. In addition, it uses the delayed gradient and the delayed information received from time-varying neighbors to update its parameters.
    \item For the analysis of the regret, we model the unbalanced graph network with delays as a row-stochastic weight matrix augmented with virtual agents and bounding both the communication and gradient delays. We prove theoretically that the proposed algorithm achieves sublinear regret, even in the presence of time-varying and non-identical communication delays and gradient delays.
    \item Our simulation result shows that the running average of each agent's local action converges over time.
\end{itemize}
The remainder of the paper is described as follows: Section II discusses the preliminaries. Section III gives the problem formulation. Section IV discusses the algorithm design. Section V gives the theoretical results. Section VI gives the simulation results. Section VII concludes the paper.

\begin{table*}[h!]
\caption{Comparison of recent works on distributed Nash equilibrium with a focus on privacy-preservation, communication delays, and feedback delays}
\centering 
    \begin{tabular}{|c| c| c|c| c|c|c|}
    \hline
\thead{paper}  & \thead{Game type} & \thead{Graph structure}  & \thead{Loss\\function} & \thead{privacy-\\preserving}& \thead{Communication\\delays} & {\thead{Feedback\\delays}} \\
\hline

\cite{zuo2023distributed,yao2025online} &aggregative &time-varying & online &no & $-$ & $-$\\

\cite{wang2024differentially,chen2025differentially,ye2021differentially} &aggregative &fixed &offline &yes & $-$ & $-$\\

\cite{chen2024communication} &non-cooperative &fixed &offline &yes & $-$ & $-$\\

\cite{lin2023statistical} &aggregative &fixed &online &yes & $-$ & $-$\\

\cite{shokri2021network}  & aggregative & fixed & offline & no & fixed/non-identical & $-$  \\

 \cite{liu2025distributed} & non-cooperative & fixed & online & no   & time-varying/non-identical & $-$  \\

\cite{zhang2020distributed} & aggregative &fixed & offline & no & time-varying/non-identical & $-$ \\

\cite{wang2022robust}  & non-cooperative & fixed & offline & no & time-varying/identical & $-$ \\

\cite{liu2023online}& aggregative & time-varying & online & no & $-$ & fixed/identical \\

 \cite{liu2024sub} & aggregative   & fixed &online & no & fixed/identical &fixed/identical \\ 

Proposed & aggregative & time-varying & online & yes & time-varying/non-identical & time-varying/non-identical\\

\hline

    \end{tabular}
    
    \label{tab:my_label}
\end{table*}

\section{Preliminaries}
\subsection{Notations} Let $\mathbb{R}^m$ and $\mathbb{R}^{m\times m}$ represent the $m-$dimensional real-valued vectors and the $m\times m$ real-valued matrices set. Let $\mathbb{0}_m$ and $\mathbb{1}_m$ represent a vector of zeros and ones, respectively. The indicator function is defined as $\mathcal{I}_\mathcal{A}(x)=  \begin{cases} 
      1 & \quad x\in\mathcal{A}, \\
      0 & \quad x\notin \mathcal{A}.
   \end{cases}$
$[\cdot]^{{T}}$ represents the Transpose of a vector or matrix. For a matrix $\mathbf{W}$, the element in its $i$-th row and $j$-th column is denoted as $[\mathbf{W}]_{ij}$. $\mathbf{W}$
is said to be row stochastic if $\mathbf{W}\mathbb{1}=\mathbb{1}$ and column stochastic if $\mathbf{W}^{{T}}\mathbb{1}=\mathbb{1}$. The inner product of two vectors is represented as $\langle \cdot \rangle$.The matrix product $\mathbf{W}(t:s):=\mathbf{W}(t)\mathbf{W}(t-1),...,\mathbf{W}(s)$ for $t\geq s\geq 0$. The 1-norm, 2-norm and dual norm of a vector $\mathbf{u}\in\mathbb{R}^m$ are denoted as $||\mathbf{u}||_1$, $||\mathbf{u}||_2$ and $||\mathbf{u}||_*:=\sup_{||\mathbf{v}||_2=1}\langle \mathbf{u}, \mathbf{v} \rangle$. For a $1$-strongly convex function $\phi:\mathbb{R}^d \rightarrow \mathbb{R}$, the projector operator $\Pi_{\mathcal{X}}^{\phi}(\mathbf{b},\eta) = \arg\min_{\mathbf{x}\in\mathcal{X}}\{\langle \mathbf{b}, \mathbf{x}\rangle + \frac{\phi(\mathbf{x})}{\eta}\}$. 
For a time-varying function $F_t(\mathbf{x}_1,\mathbf{x}_2)$, its gradient is represented as $\triangledown F_t(\mathbf{x}_1,\mathbf{x}_2)$ and $\triangledown_2 F_t(\mathbf{x}_1,\mathbf{x}_2)$ with respect to $\mathbf{x}_1$ and $\mathbf{x}_2$ respectively.
Bregman divergence $B_{\phi}(\mathbf{x},\mathbf{y})=\phi(\mathbf{x})-\phi(\mathbf{y})-\langle \triangledown \phi(\mathbf{y}),\mathbf{x}-\mathbf{y}\rangle$.

\subsection{Graph Theory}
A time-varying unbalanced graph network is represented as $\mathcal{G}(t)=(\mathcal{V},\mathcal{E}(t))$, where $\mathcal{V}=\{1,...,V\}$ is the set of agents in the graph network and $\mathcal{E}(t)\subseteq \mathcal{V}\times \mathcal{V}$ is the time-varying edge set. We define the in-neighborhood of an agent $i$ at time $t$ as $\mathcal{N}_{i,t}^{\text{in}}:= \{j\in\mathcal{V}:(i,j)\in\mathcal{E}(t)\}$, and the out-neighborhood of the agent $i$ at time $t$ as $\mathcal{N}_{i,t}^{\text{out}}:= \{j\in\mathcal{V}:(j,i)\in\mathcal{E}(t)\}$.  The in-degree of agent $i$ is denoted as $d_{i,t}=|\mathcal{N}^{in}_{i,t}|$. The time-varying weight matrix is denoted as $\mathbf{W}(t)$ and $[\mathbf{W}(t)]_{ij}=\frac{1}{d_{i,t}}$ when $j\in\mathcal{N}_{i,t}^{\text{in}}$, and $[\mathbf{W}(t)]_{ij}=0$ otherwise. 
The graph is said to be unbalanced if there is at least an agent $i\in\mathcal{V}$ such that $|\mathcal{N}_{i,t}^{\text{in}}|\neq |\mathcal{N}_{i,t}^{\text{out}}|$.

\section{Problem Formulation}
Consider a non-cooperative game in an online setting denoted as $\mathbb{G}(t)=\{\mathcal{V},\Omega, F(t)\}$ where $\mathcal{V}=(1,...,V)$ represents the set of agents, $\Omega = \Omega_1 \times \Omega_2 \times ... \times \Omega_V$ represents the joint action of the agents with $\Omega_i\subseteq \mathbb{R}^m$, and $F_t =\{F_{1,t},F_{2,t},...,F_{V,t}\}$ represents the set of time-varying cost functions with $F_{i,t}:\Omega_i\rightarrow \mathbb{R}$. The action profile is defined as $\mathbf{x}(t) =col(\mathbf{x}_1(t),...,\mathbf{x}_V(t)):=(\mathbf{x}_i(t),\mathbf{x}_{-i}(t))$ with $\mathbf{x}_i(t)\in\Omega_i$ and $\mathbf{x}_{-i}(t):=col(\mathbf{x}_1(t),...,\mathbf{x}_{i-1}(t), \mathbf{x}_{i+1}(t),...,\mathbf{x}_{V}(t))$ representing the actions of other agents excluding agent $i$. Due to the online nature of the game, at time $t$, the cost function $F_{i,t}(\mathbf{x}_{i}(t),\mathbf{x}_{-i}(t))$ depends on the aggregate function $\Psi(\mathbf{x}(t)):=\frac{1}{V}\sum_{j=1}^V \psi_j(\mathbf{x}_j(t))$ where $\psi_j(\mathbf{x}_j(t)):\mathbb{R}^m \rightarrow \mathbb{R}^m$, and henceforth denote $F_{i,t}(\mathbf{x}_i(t),\Psi(\mathbf{x}(t)))$ as agent $i$'s
cost function. The agent $i$'s learning objective at each time $t\in\{1,...,T\}$ is to solve the optimization problem below:
\begin{equation}
    \mathbf{x}^*_i(t) = \arg
    \min_{\mathbf{x}_i(t)\in\Omega_i}
    F_{i,t}(\mathbf{x}_i(t),\Psi(\mathbf{x}_i(t),\mathbf{x}^*_{-i}(t))).
    \label{eqn 1}
\end{equation}
In a distributed setting, no agent has access to the aggregate function $\Psi(\mathbf{x}(t))$ due to the absence of a central coordinator. However, agents can estimate the aggregate function by communicating with their neighbors in a time-varying graph network. The goal is to design an online learning algorithm that learns the optimal solution to Problem (\ref{eqn 1}), referred to as the time-varying Nash equilibrium, with privacy guarantees and delay tolerance. 

\textbf{Definition 1 \textcolor{blue}{\cite{lin2023statistical,deng2023distributed}}:} The action profile $\mathbf{x}^*(t)=(\mathbf{x}_1^*(t),...,\mathbf{x}_V^*(t))$ is the time-varying Nash equilibrium of (\ref{eqn 1}) if for any $i\in\mathcal{V}$ and any time $t\in\{1,...,T\}$
\begin{equation}
\begin{split}
   F_{i,t}(\mathbf{x}^*_i(t),\Psi(\mathbf{x}^*(t)))\\
   \leq 
   F_{i,t}\bigg(\mathbf{x}_{i}(t),\frac{1}{V}\sum_{j=1,j\neq i}^V \psi_j(\mathbf{x}^*_j(t)) + \frac{1}{V}\psi_i(\mathbf{x}_i(t))\bigg)\\ = F_{i,t}(\mathbf{x}_i(t), \Psi(\mathbf{x}_i(t), \mathbf{x}^*_{-i}(t))).
   \end{split}
\end{equation}
The time-varying Nash equilibrium depicts that no agent can unilaterally adjust its action to reduce its cost value at any time $t$.

The gradient of agent $i\in\mathcal{V}$ in a delay-free and non-private environment is given by 
\begin{equation}
\begin{split}
 \triangledown F_{i,t}(\mathbf{x}_i(t), \Psi(\mathbf{x}(t)))=\triangledown_1 F_{i,t}(\mathbf{x}_i(t),\Psi(\mathbf{x}(t))) +\\ \frac{\triangledown \psi_i(\mathbf{x}_i(t))}{V}\triangledown_2 F_{i,t}(\mathbf{x}_i(t),\Psi(\mathbf{x}(t))), 
 \end{split}
\end{equation}
And the pseudogradient is given as
\begin{equation}
\begin{split}
\triangledown F_t(\mathbf{x}(t),\Psi(\mathbf{x}(t))) =\\  col(\triangledown F_{1,t}(\mathbf{x}_1(t), \Psi(\mathbf{x}(t))),...,\triangledown F_{V,t}(\mathbf{x}_V(t),\Psi(\mathbf{x}(t))))
\end{split}
\end{equation}
\subsection{Necessary Assumptions}
\textbf{Assumption  1:} The time-varying unbalanced graph network is $B$-strongly connected, i.e., there exists a constant $B>0$ such that the edge set  $\bigcup_{t=kB}^{(k+1)B-1}$ is strongly connected for any $k\geq 0$.

\textbf{Remark 1:} Assumption 1 is less stringent compared to allowing each $\mathbb{G}(t)$ to be time-varying. Assumption 1 is commonly used for time-varying graphs in decentralized learning \cite{chen2021communication,saadatniaki2020decentralized}.


\textbf{Assumption 2:} For any agent $i\in\mathcal{V}$, the convex set $\Omega_i\subseteq\mathbb{R}^m$ is non-empty and compact, the cost function $F_{i,t}(\mathbf{x}_i(t),\Psi(\mathbf{x}(t)))$ is convex and continuously differentiable in $(\mathbf{x}_i(t), \Psi(\mathbf{x}(t)))$, and $\psi_i(\mathbf{x}_i(t))$ is continuously differentiable in $\mathbf{x}_i(t)$.


\textbf{Assumption 3:} The pseudo-gradient $\triangledown F_t(\mathbf{x}_t,\Psi(\mathbf{x}(t)))$ is $\mu-$strongly monotone for some $\mu>0$.

A unique Nash equilibrium exists based on Assumptions 2 and 3 at each time $t$, such that $\triangledown F_t(\mathbf{x}^*(t),\Psi(\mathbf{x}^*(t))) = \mathbb{0}_{m\times N}$. 

\textbf{Assumption 4:} (a) There exists constant $L>0$, such that $|F_{i,t}(\mathbf{x}_i, \Psi(\mathbf{x}_i,\mathbf{x}_{-i}))- F_{i,t}(\mathbf{y}_i, \Psi(\mathbf{y}_i, \mathbf{x}_{-i}))|\leq L ||\mathbf{x}_i-\mathbf{y}_i||$. This implies $||\triangledown_1 F_{i,t}(\mathbf{x}_i,\Psi(\mathbf{x}))||\leq L$.

(b) $\triangledown_1 F_i(\mathbf{x}_i,\Psi(\mathbf{x}))$ is $G-$Lipschitz continuous, namely $F_{i,t}(\mathbf{y}_i,\Psi(\mathbf{y},\mathbf{x}_{-i}))\leq F_{i,t}(\mathbf{x}_{i},\Psi(\mathbf{x})) + \langle\triangledown_1 F_{i,t}(\mathbf{x}_i,\Psi(\mathbf{x})), \mathbf{y}_i-\mathbf{x}_i\rangle + \frac{G}{2}||\mathbf{x}_i-\mathbf{y}_i||^2$ holds for $\mathbf{x}_i, \mathbf{y}_i\in\mathcal{X}$. 

\textcolor{blue}{\textbf{Remark 2:} The bounded constraint assumption in Assumption 2 is necessary to ensure that the projection operator $\Pi_\mathcal{X}^\phi$ is unique. Assumptions 2 and 3 are necessary to ensure the uniqueness of the time-varying Nash equilibrium. Assumption 4 is necessary to obtain a sublinear regret bound and also to bound sensitivity in privacy analysis. Other similar works have used these assumptions, such as \cite{chen2024communication,nguyen2025distributed,ye2021differentially,lu2023privacy,xiong2020privacy}.}

\textbf{Assumption 5:} For $\mathbf{x}_i^*(t)\in \arg\min_{\mathbf{x}_i(t)\in\mathcal{X}}F(\mathbf{x}_i(t),\Psi(\mathbf{x}_i(t), \mathbf{x}^*_{-i}(t)))$, there always exists a constant $R>0$, such that $\phi(\mathbf{x}^*_i(t))\leq \frac{R^2}{2}$ and $B_{\phi}(\mathbf{x}^*_i(t),\mathbf{x}_i(t))\leq R^2$ for any $\mathbf{x}_i(t)\in\mathcal{X}$.

\textbf{Remark 3:} The Bregman's divergence generalizes the Euclidean square distance. Specifically, when $\phi(\mathbf{x}_i(t))=||\mathbf{x}_i(t)||^2$, we obtain the Euclidean square distance, and Assumption 5 implies $||\mathbf{x}_i(t)-\mathbf{x}^*_i(t)||^2 \leq R^2$, for any $\mathbf{x}_i(t)\in\mathcal{X}$.


In online learning for aggregative games, the dynamic regret is the performance metric to evaluate any proposed algorithm. The dynamic regret  for agent $i\in \mathcal{V}$ is defined as
\begin{equation}
\begin{split}
R_i(T) = \sum_{t=1}^T \sum_{i=1}^V F_{i,t}(\mathbf{x}_i(t),\Psi(\mathbf{x}_i(t),\mathbf{x}^*_{-i}(t)))-\\ \sum_{t=1}^T\sum_{i=1}^V F_{i,t}(\mathbf{x}^*_i(t),\Psi(\mathbf{x}^*(t))).   
\end{split}
\end{equation}
A good online algorithm must achieve a sublinear regret such that $\lim_{T\rightarrow\infty} \frac{R_i(T)}{T}=0$.

\subsection{Differential Privacy}
Differential privacy is used to protect each agent's actions shared with neighbors in the time-varying graph network. Differential privacy involves the addition of Laplace noise drawn from a Laplacian distribution to the agents' actions.  We define the concept of adjacency for the cost functions in the online setting for each agent $i\in\mathcal{V}$ over the time horizon as follows:

\textbf{Definition 2:} Two sets of cost functions are said to be adjacent if ${Z}_{i, T}:=\{F_{i,1},...F_{i, T}\}$ and ${Z}^\prime_{i,T}:=\{F^\prime_{i,1},...,F^\prime_{i,T}\}$  differs in just only one cost function, i.e., $F_{i,j}\neq F^\prime_{i,j}$ but $F_{i,k}=F^\prime_{i,k}$ for all $k\in\{1,...,T\}$ and $j\neq k$. 

Definition 2 infers that if two sequences of cost functions differ in only one cost function, then the two sequences are adjacent. Leveraging Definition 2, we now define differential privacy as follows:

\textbf{Definition 3:} Consider two adjacent cost functions $Z_{i,T}$ and $Z^\prime_{i,T}$. If 
\begin{equation}
   \mathbb{P}[\mathcal{A}(Z_{i,T})\in S]\leq  \exp(\epsilon)\cdot\mathbb{P}[\mathcal{A}(Z^\prime_{i,T})\in S],
\end{equation}
then the randomized algorithm $\mathcal{A}$ is said to be $\epsilon-$differentially private, where $S$ is a subset of the range of $\mathcal{A}$ and $\epsilon$ is the privacy parameter. The $\epsilon-$differential privacy is also referred to as ``pure differential privacy", which is obtained when noise is drawn from a Laplacian distribution. 

The above definition of $\epsilon$-differential privacy guarantees that even if an adversary has access to all time-varying cost functions of an agent within the graph network, it is unlikely to infer any action of the agent with high confidence. Moreover, a smaller value of $\epsilon$ indicates a stronger level of privacy protection. It is also important to highlight that the $\epsilon$-differential privacy is more rigorous than other relaxed (approximate) variants, such as ($\epsilon$, $\delta$)-differential privacy and zero-concentrated differential privacy.

\textbf{Definition 4:} The sensitivity at time $t>1$ is defined as $\Delta_t=\sup_{Z_{i,t}, Z_{i,t}^\prime: \text{Adj}(Z_{i,t},Z^\prime_{i,t})}||\mathcal{A}(Z_{i,t})-\mathcal{A}(Z^\prime_{i,t})||_1$.

The sensitivity is used to regulate the amount of noise to be added to guarantee differential privacy.

\subsection{Time-Varying and Non-Identical Communication and Feedback Delays }
We consider the situation where there are communication delays in sharing information with other agents in the neighborhood. We also consider gradient feedback delays. We describe the delay model as follows:

For communication delays, the information sent by agent $j$ at time $t$  to agent $i$ is delayed by $\tau_{ij}(t)$. So, agent $i$ receives this information at time $t+\tau_{ij}(t)$. For feedback delays, the gradient of the cost function of agent $i$ at time $t-\tau_i(t)$ is used at time $t$.

The following assumptions hold for both communication and feedback delays:

\textbf{Assumption 6:} The communication delay $\tau_{ij}(t)$ between any two agents $i$ and $j$ is bounded between the interval $[0,\tilde{\tau}]$, i.e., $0\leq\tau_{ij}(t)\leq \tilde{\tau}<\infty$. Furthermore, there is no communication delay with each agent accessing its own parameters because $\tau_{ii}(t)=0$.

\textbf{Assumption 7:} The feedback delay for any agent $i$ at any time $t$ is bounded by $\tilde{\tau}$, i.e., $0\leq \tau_{i}(t)\leq \tilde{\tau}<\infty$.

Note that, unlike in existing works \cite{liu2024sub,liu2023online}, the communication and feedback delays are time-varying and non-identical across time and agents, although they are bounded.

\section{Algorithm Design}
To solve the non-cooperative game problem over a time-varying graph network in a distributed manner, we employ online distributed dual averaging \cite{hosseini2013online,lu2023privacy}. First, we propose the algorithm without considerations for privacy or delays. Then, we extend the proposed algorithm to the private and delay-tolerant setting. We propose the algorithm as follows:

\begin{equation}
 \mathbf{b}_i(t+1) = \sum_{j=1}^V [\mathbf{W}(t)]_{ij} \mathbf{b}_{j}(t) +\frac{\mathbf{g}_{i,t}(\mathbf{x}_i(t), \mathbf{v}_i(t))}{{y}_{ii}(t)},   
\end{equation}

\begin{equation}
 \mathbf{y}_i(t+1) =\sum_{j=1}^V [\mathbf{W}(t)]_{ij}\mathbf{y}_j(t),   
\end{equation}

\begin{equation}
 \mathbf{x}_i(t+1)=   \Pi^{\phi}_{\mathcal{X}} (\mathbf{b}_i(t+1),\eta_{t+1}),
\end{equation}

\begin{equation}
  \hat{\mathbf{x}}_i(t+1) = \frac{t}{t+1}\hat{\mathbf{x}}_i(t) + \frac{1}{t+1}\mathbf{x}_i(t+1),  
\end{equation}

\begin{equation}
  \mathbf{v}_i(t+1) = \sum_{i=1}^V [\mathbf{W}(t)]_{ij}\mathbf{v}_j(t) + \psi_i(\hat{\mathbf{x}}_i(t+1))-\psi_i(\hat{\mathbf{x}}_i(t)),  
\end{equation}
where the dual variable $\mathbf{b}_i(0)=\mathbb{0}_m$, $\triangledown F_{i,t} $ is the gradient of the agent $i$ abbreviated as $\mathbf{g}_{i,t}$. Because in a distributed setting, no agent has access to the aggregate function, we estimate $\Psi(\mathbf{x}(t))$ with $\mathbf{v}_i(t)$ for each agent using information received from neighbors. Let $\mathbf{v}_i(0)=\psi_i(\mathbf{x}_i(0))$. To compensate for the unbalancedness of the graph, each agent estimates the left eigenvector of the weight matrix that corresponds to an eigenvalue of $1$ as $\mathbf{y}_{i}(t)$ by aggregating information from neighbors. This is achieved by first indexing the agents, such that $\mathbf{y}_i(0)= [0,...,1,...,0]^{\mathbb{T}}\in\mathbb{R}^V$ where $1$ is in the $i$th entry and $0$ everywhere else. \textcolor{blue}{It is to be noted that $y_{ii}(t)$ cannot be zero for all $t\geq 0$, as discussed in \cite{mai2016distributed, priolo2013decentralized}}. The operation $\Pi^{\phi}_{\mathcal{X}}$ generates $\mathbf{x}_i(t+1)$ while $\hat{\mathbf{x}}_i(t+1)$ is the running local average of $\mathbf{x}_i(t+1)$. The running local average allows each agent to access the running average without storing previous iteration results. Let $\mathbf{x}_i(0)=\hat{\mathbf{x}}_i(0)\in\mathbb{R}^m$. The non-diminishing step size is $\eta_{t+1}=\frac{\gamma}{\sqrt{t+1}}$ with $\gamma>0$ and $t\geq 0$. The proposed algorithm assumes a row-stochastic graph network where agent $i$ does not need to know its out-degree but only needs to broadcast its private information. In addition, agent $i$ can independently assign weights to the information it receives from its in-neighbors. Our choice of row stochasticity is based on existing work that shows that algorithms designed for column stochasticity suffer in the presence of communication failure due to agents' inability to adjust the outgoing weights of communication links to out-neighbors \cite{huang2025dual,xi2018linear,zhang2017distributed}. 

We now propose a new distributed dual averaging algorithm that is both private and resilient to time-varying and non-identical communication and feedback delays, as outlined in Algorithm 1 below.
\begin{algorithm}
\caption{Online Distributed Dual Averaging Algorithm for Agent $i$ With Privacy and Delay Tolerance}
\begin{algorithmic}[1]
\State \textbf{Initialization}: Time duration $T$, step size $\eta_{t+1} = \frac{\gamma}{\sqrt{t+1}}$, with $\gamma>0$, $\mathbf{x}_i(0)\in\mathbb{R}^m$, $\mathbf{v}_i(0)=\psi_i(\mathbf{x}_i(0))$.
\State \textbf{Output:} $\hat{x}_i(T)$.
\For{$t = 0, ... ,T$}
\State Generate Laplace noise $\mathbf{n}_{i}(t)\sim \mathcal{L}(\mathbf{0},\sigma_t\mathbb{I}_m)$.
\State $\tilde{\mathbf{v}}_i(t)$ = $\mathbf{v}_i(t)$ +  $\mathbf{n}_{i}(t)$ and $\tilde{\mathbf{b}}_{i}(t)$ = $\mathbf{b}_{i}(t) +\mathbf{n}_{i}(t)$.
\State Send $[\mathbf{W}(t)]_{ji}\tilde{\mathbf{b}}_i(t)$ and $[\mathbf{W}(t)]_{ji}\tilde{\mathbf{v}}_i(t)$  
to out-neighbors $j\in\mathcal{N}_{i,t}^{\text{out}}$ and receive from in-neighbors $j\in \mathcal{N}_{i,t}^{\text{in}}$ the sets $\{[\mathbf{W}(t-\tau_{ij}(t))]_{ij}\tilde{\mathbf{b}}_j(t-\tau_{ij}(t))\}$, and $\{[\mathbf{W}(t-\tau_{ij}(t))]_{ij}\bar{\mathbf{v}}_j(t-\tau_{ij}(t))\}$ that arrive at time $t$ where $\tau_{ij}(t)\leq \tilde{\tau}$.
\State $\mathbf{b}_i(t+1) = [\mathbf{W}(t)]_{ii}{\mathbf{b}}_i(t) + \sum_{j\in\mathcal{N}_{i,t}^{\text{in}}}\sum_{r=0}^{\tilde{\tau}}[\mathbf{W}(t-r)]_{ij}\tilde{\mathbf{b}}_j(t-r)\mathcal{I}_{t-r,ij}(r)+\frac{\mathbf{g}_{i,t-\tau_{i}(t)}(\mathbf{x}_i(t-\tau_{i}(t)),\mathbf{v}_i(t-\tau_{i}(t)))}{{y}_{ii}(t)}$ \newline with $\mathcal{I}_{t,ij}(\tau)= \begin{cases} 
      1 & \quad \tau_{ij}(t)=\tau, \\
      0 & \quad \text{otherwise.}
   \end{cases}$

\State $\mathbf{y}_i(t+1)=\sum_{j=1}^V [\mathbf{W}(t)]_{ij}\mathbf{y}_j(t)$. 
\State $\mathbf{x}_i(t+1)=\Pi^{\phi}_{\mathcal{X}}(\mathbf{b}_i(t+1),\eta_{t+1})$.
\State $\hat{\mathbf{x}}_i(t+1)=\frac{t}{t+1}\hat{\mathbf{x}}_i(t) + \frac{1}{t+1}\mathbf{x}_i(t+1)$.
\State $\mathbf{v}_i(t+1) =  [\mathbf{W}(t)]_{ii}{\mathbf{v}}_i(t) + \sum_{j\in\mathcal{N}_{i,t}^{\text{in}}}\sum_{r=0}^{\tilde{\tau}}[\mathbf{W}(t-r)]_{ij}\tilde{\mathbf{v}}_j(t)\mathcal{I}_{t-r,ij}(r)+ \psi_i(\hat{\mathbf{x}}_i(t+1))-\psi_i(\hat{\mathbf{x}}_i(t))$
\EndFor
\end{algorithmic}
\end{algorithm}
\textcolor{blue}{The operation of Algorithm 1 is discussed as follows: Step 1 is the initialization of parameters. Step 2 is the output we seek at the end of the iteration. Step 3 shows the algorithm's iteration over the time horizon $T$.} In Step 4, Laplace noise is drawn from a Laplacian distribution with a time-varying parameter $\sigma_t$. \textcolor{blue}{The Laplace noise will be used to randomize the sensitive parameters.} In Step 5, Laplace noise is added to \textcolor{blue}{sensitive parameters}, which are the estimated aggregate $\mathbf{v}_i(t)$ and the dual variable $\mathbf{b}_i(t)$ to randomize them for privacy protection. The private version is denoted as $\tilde{\mathbf{b}}_i(t)$ and $\tilde{\mathbf{v}}_i(t)$. In Step 6, agent $i$ sends $[\mathbf{W}(t)]_{ji}\tilde{\mathbf{b}}_i(t)$, $[\mathbf{W}(t)]_{ji}\tilde{\mathbf{v}}_i(t)$ and $[\mathbf{W}(t)]_{ji}\mathbf{y}_i(t)$ to its out-neighbors and receives private information sent in previous times but arrives at time $t$ from its in-neighbors, that is, the sets $\{[\mathbf{W}(t-\tau_{ij}(t))]_{ij}\tilde{\mathbf{b}}_j(t-\tau_{ij}(t))\}$, and $\{[\mathbf{W}(t-\tau_{ij}(t))]_{ij}\tilde{\mathbf{v}}_j(t-\tau_{ij}(t))\}$. \textcolor{blue}{Note that the received private information from the in-neighbors experiences random communication delays}. 
\textcolor{blue}{In step 7, the agent $i$ updates the parameter $\mathbf{b}_i(t+1)$ using its weighted private information, the weighted sum of the received private information from its in-neighbors in a consensus fashion, and its compensated gradient feedback. The compensation accounts for the unbalancedness of the time-varying graph. However, the received private information from its in-neighbors experiences random communication delays, while its compensated gradient feedback also experiences a random delay.} The gradient feedback $\mathbf{g}_{i,t-\tau_{i}}(\mathbf{x}_i(t),\mathbf{v}_i(t))$ is randomly delayed by $\tau_{i}(t)\leq \tilde{\tau}<\infty$. The indicator function $\mathcal{I}_{t,ij}(\tau)= \begin{cases} 
      1 & \quad \tau_{ij}(t)=\tau, \\
      0 & \quad \text{otherwise.}
   \end{cases}$ indicates that the information from the in-neighbors reaches the agent $i$ after a delay. Note that there is no communication delay for an agent to access its own information, nor does an agent need to randomize its parameter with Laplace noise before it accesses it. Hence, $[\mathbf{W}(t)]_{ii}\mathbf{b}_i(t)$ does not have a delay term or uses a private version of the dual variable. \textcolor{blue}{Summing the received private information from the in-neighbors over $\tilde{\tau}$ multiplied by the indicator function in the second term of Step 7, is equivalent to summing only the private information from neighbors that arrive at time $t$.}
    Step 8 also computes an estimate of the eigenvector in a distributed fashion. Since both the index number of every agent and the weight matrix are common knowledge, agent $i$ can compute $\mathbf{y}_{i,t+1}$ for all agents without requesting them to be sent, thus removing communication delay during the estimation of the eigenvector. \textcolor{blue}{For instance, the index for any agent $i$ at time $t=0$ is $\mathbf{y}_i(0)=[0,...,1,...,0]^\mathbb{T}\in\mathbb{R}^V$ where $1$ is in the $i$th position and $0$ elsewhere. This is public information. Moreover, the time-varying weight matrix $\mathbf{W}(t)$ for all $t\geq 0$ is public information. Thus, updating $\mathbf{y}_i(t+1)$ does not require the in-neighbors of agent $i$ to send $[\mathbf{W}(t)]_{ij}$ or $\mathbf{y}_j(t)$}. This is unlike in \cite{wang2023distributed}, where communication delays occur because the in-neighbors need to send parameters for update. Moreover, in step 8, each agent does not need to know the out-degree of its incoming neighbors to remove unbalancedness as required in the push-sum approach in \cite{wang2023distributed}. Step 9 updates agent $i$'s
action. Step 10 computes the running local average. \textcolor{blue}{The running local average allows agent $i$ to access the running average without storing previous iteration results.}  Step 11 updates the aggregate estimates with privacy considerations and delay robustness. \textcolor{blue}{Again, in the second term of step 11, summing the received private information from in-neighbors over $\tilde{\tau}$ is equivalent to summing only private information from in-neighbors that arrives at time $t$.} \textcolor{blue} {Note that step $11$ is necessary because the agent $i$ cannot observe the actions of every agent to compute the aggregate function. Therefore, it estimates it by using only parameters from its in-neighbors in the time-varying graph network. }

The proposed Algorithm 1 enables each agent to update its parameters without waiting for information to arrive, i.e., update without missing rounds. Delayed information received at a given time is used in the update process at that time. Also, agents do not have to wait for the gradient at the current time before updating. 
\textcolor{blue}{For analysis}, we can model the graph network with communication delay as a row-stochastic weight matrix by adding $\tilde{\tau}V$ virtual agents in the network \cite{wang2023distributed,huang2025dual}. Thus, the total number of agents become $V^\prime = V + \tilde{\tau}V$  and the set of all agents is represented as $\mathcal{V}^\prime = \{1,...,V^\prime\}$. We represent the matrix $\mathbf{W}^\prime(t)\in \mathbb{R}^{V^\prime \times V^\prime}$ as follows

\[
\mathbf{W}^\prime(t) = \begin{bmatrix} 
    \mathbf{W}^0(t) & \mathbf{W}^1(t) & \dots  &\mathbf{W}^{\tilde{\tau}-1}(t) & \mathbf{W}^{\tilde{\tau}}(t)\\
    \mathbb{I}_V & 0 & \dots & 0 & 0\\
    0 & \mathbb{I}_V & \dots & 0 & 0\\
    \vdots & \vdots & \ddots & \vdots &\vdots\\
    0 & 0  & \dots &  \mathbb{I}_V & 0
    \end{bmatrix}
\]

where $[\mathbf{W}^r(t)]_{ij} = \begin{cases} 
      [\mathbf{W}(t)]_{ij} & \quad \tau_{ij}(t)=r, \\
      0 & \quad \text{otherwise.}\end{cases}$.
Hence, any non-identical and time-varying communication delays less than the maximum delay $\tilde{\tau}$ can be represented by the augmented weight matrix $\mathbf{W}^\prime (t)$, which is also a row stochastic matrix because the original matrix $\mathbf{W}(t)$ is row-stochastic \cite{hadjicostis2013average}. \textcolor{blue}{Note that the augmented weight matrix is used only for analysis and not during the operation of the proposed algorithm.}

\section{Theoretical Results}
\subsection{Regret Analysis}
\textbf{Lemma 1 \cite{xie2018distributed, xiong2022distributed,huang2025dual}:} Suppose Assumption 1 holds, then, for any $t\geq 0$, a normalized vector ${\pi}(t)$ exists, such that ${\pi}(t) = \{\pi_i(t),...,\pi_{V^\prime}(t)\}^{\mathbb{T}}$ and $\mathbb{1}^{\mathbb{T}}\pi(t)=1$, and 

a) For any $i,j\in\mathcal{V}^\prime$ and $t\geq s\geq 0$, $|[\mathbf{W}^{\prime}(t:s)]_{ij}-\pi_j(s)|\leq C\lambda^{t-s}$ with $C>0$ and $\lambda\in(0,1)$, while $\frac{1}{\theta}\leq y_{ii}(t)\leq 1$ $\forall$ $i,j\in\mathcal{V}^\prime$, $t\geq 0$ with $\theta$ a constant.

b) There exists a constant $0\leq \rho \leq 1$, and $\beta \geq \rho ^{(V^\prime -1)B}$ such that $\beta \leq \pi_i(t)\leq 1$ for any $i\in \mathcal{V}^\prime$ and $t\geq 0$. Note that $B$ is from Assumption 1.

c) For any $t\geq 0$, $\pi^{\mathbb{T}}(t+1)\mathbf{W}^\prime(t)=\pi^{\mathbb{T}}(t)$.

\textbf{Lemma 2 \cite{wang2014cooperative}:} Let $\mathbf{x}^+$ be the minimizer of  $\langle \mathbf{k},\mathbf{x}\rangle + h\phi(\mathbf{x})$ for all $\mathbf{x}\in\mathcal{X}$. Then, for any $\mathbf{x}\in\mathbb{R}^m$,  $\langle \mathbf{k},\mathbf{x}\rangle + h\phi(\mathbf{x})\geq \langle \mathbf{k},\mathbf{x}^+\rangle + h\phi(\mathbf{x}^+)+ hB_{\phi}(\mathbf{x},\mathbf{x}^+)$.

\textbf{Lemma 3:} Define $\bar{\mathbf{b}}(t) = \sum_{i=1}^{V} \pi_i \mathbf{{b}}_i(t)$. Then, $\bar{\mathbf{b}}(t+1)= \bar{\mathbf{b}}(t)+ \sum_{i=1}^{V}\pi_i(t+1)\frac{g_{i,t-\tau_i(t)}}{y_{ii}(t)}$.

\textbf{Proof:} Refer to Appendix A.

\textbf{Lemma 4:}  Define $\mathbf{z}(t) = \Pi_{\mathcal{X}}^\phi(\bar{\mathbf{b}}(t),\eta_t)$. For any $\mathbf{x}^*$, it holds that $\sum_{t=1}^T \sum_{i=1}^{V}\pi_i \langle \frac{\mathbf{g}_{i,t-\tau_i(t)-1}}{y_{ii}(t-1)}, \mathbf{z}(t)-\mathbf{x}^*_i(t)\rangle \leq \frac{\phi(\mathbf{x}^*_i(T))}{\eta_T}$.

\textbf{Proof:} Refer to Appendix B.

\textbf{Lemma 5 \cite{duchi2011dual}:} For any vector $\mathbf{u}$,$\mathbf{v}\in\mathbb{R}^m$, we have
\begin{equation}
 ||\Pi^{\phi}(\mathbf{u},\eta)-\Pi^{\phi}(\mathbf{v},\eta)||\leq \eta||\mathbf{u}-\mathbf{v}||_*.   
\end{equation}

\textbf{Theorem 1:} \textcolor{blue}{While Assumptions 1 - 5 hold}, the dynamic regret bound for the proposed Algorithm 1 with $\eta_t = \frac{\gamma}{\sqrt{t+1}}$ is given as 
\begin{equation}
\begin{split}
   \frac{1}{V} \sum_{i=1}^T \sum_{i=1}^{V}F_{i,t}(\mathbf{x}_i(t),\Psi(\mathbf{x}_i(t),\mathbf{x}^*_{-i}(t))-\\
   \frac{1}{V}\sum_{i=1}^T \sum_{i=1}^{V} F_{i,t}(\mathbf{x}_i^*(t),\Psi(\mathbf{x}^*(t)))\\ \leq O\bigg(\frac{\tilde{\tau}^2 V\sigma_{max}}{(1-\lambda)\sqrt{T}}\bigg).
   \end{split}
   \label{eqn 12b}
\end{equation}
where $\sigma_{max} :=\max_{0\leq t\leq T}\sigma(t)$.

\textbf{Proof:} Refer to Appendix C.

\textbf{Remark 4:} \textcolor{blue}{The regret bound in (\ref{eqn 12b}) shows the coupling effect of privacy and delay. Both privacy and delay directly increase the regret bound. Thus, carefully adjusting the delay bound $\tilde{\tau}$ and the maximum noise variance $\sigma_{max}$ is necessary to ensure a sublinear regret.}

\subsection{Privacy Analysis}
\textbf{Lemma 6:} While Assumptions 1 and 2 hold, the sensitivity $\Delta_t$ is bounded as
\begin{equation}
    \Delta_t \leq 2L\theta\sqrt{m},
\end{equation}
where $\theta$ is gotten from Lemma 1 and $m$ is the dimension of $\mathbf{b}_i, i\in\mathcal{V}$.

\textbf{Proof:} Refer to Appendix D

\textbf{Theorem 2:} The Laplace parameter $\sigma_t = \frac{\Delta_t}{\epsilon_t}$. Under Assumptions 1 and 2, Algorithm 1 achieves $\epsilon_k-$differential privacy at each time $t$. Over the time horizon, Algorithm 1 achieves $\hat{\epsilon}$-differential privacy, where $\hat{\epsilon}=\sum_{t=1}^T (\Delta_t/\sigma_t)$.

\textbf{Proof:} Refer to Appendix E.

\section{Simulation Results}
For our simulation, we consider the online Nash-Cournot game with communication delay $\tau_{ij}(t)$ and feedback delay $\tau_i(t)$. We consider five firms (agents) contained in the set $\mathcal{V} =\{1,...,5\}$ in a time-varying unbalanced graph network $\mathcal{G}=(\mathcal{V},\mathcal{E}(t))$ with time-varying row-stochastic adjacent matrix $\mathbf{W}(t)$ as shown in Fig. 1. The edge weight for edge $(i, j)\in\mathcal{E}(t)$ is given as $[\mathbf{W}(t)]_{ij}=\frac{1}{d_{i,t}}$, and there is a self-loop for all firms. In Fig. 1, unreliable communication occurs between the firm $2$ and the firm $4$ only in odd numbers of iterations \cite{huang2025dual}. The firm $2$ is unaware of the communication delay. All firms are assumed to undergo time-varying feedback delays, which cause gradient delays.  

\begin{figure}[h!]
    \centerline{\includegraphics[width=0.7\linewidth]{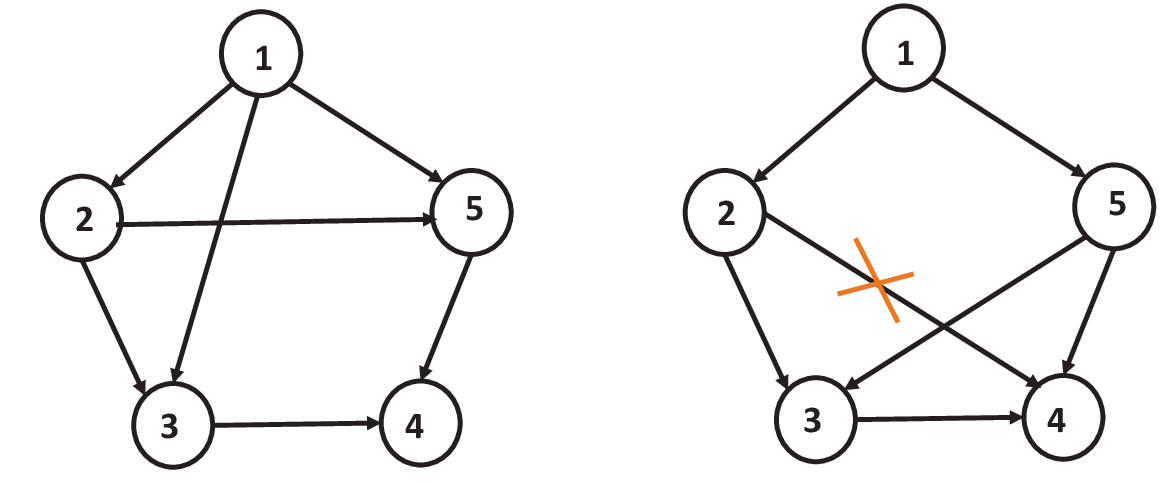}}
    \caption{Time-varying unbalanced graph networks with a broken path at odd iterations.}
    \label{fig1}
\end{figure}

In the Nash-Cournot game, the quantity of products produced by a firm $i$ is represented as $\mathbf{x}_i(t)\in\mathbb{R}^m$. We let $m=1$ for simplicity. The production price of the firm $i$ at time $t$ is given as $p_{i,t} = 4(i+1)\sin(t/6)+50i$. The market price of firm $i$ at time $t$ is $m_{i,t}= 850 - 10\sin(t/6)-\sum_{j=1}^5 {x}_{j}(t)$. The time-varying cost function of firm $i$ is given as $F_{i,t}({x}_i(t),\Psi({x})) = (p_{i,t}-m_{i,t}){x}_{i}(t)$. The aggregation function to be estimated by the firms is $\sum_{j=1}^5 x_j(t)$. To ensure that the time-varying Nash equilibrium is constrained within the action sets, we set the local action sets of the firms as $\Omega_1 = [-5, 5]$, $\Omega_2 = [0, 10]$, $\Omega_3 = [-4, 8]$, $\Omega_4 = [3, 12]$, and $\Omega_5 = [-1, 6]$, respectively. The step size is $\eta_t = \frac{\gamma}{\sqrt{t+1}}$ with $\gamma = 1$, and the $1-$strongly convex function chosen for firm $i$ is $\phi_i(x_i(t))= \frac{1}{2}|x_i(t)|^2$. We initialize $x_1(0) = -1$, $x_2(0)=2$, $x_3(0)= 2$, $x_4(0)=5$, and $x_5(0)= 1$. \textcolor{blue}{We randomize the parameters with Laplace noise as shown in Algorithm 1, we let the sensitivity $\Delta_t \leq 1$ for all $t\in \{0,..,T\}$ and choose $\epsilon = 0.2$.
\textit{}We run simulations to determine the effects of learning rate, privacy, communication delays, and feedback delays. We track the trajectory of the agents' actions. We also track the average loss \cite{wu2022decentralized} and observe that it converges toward zero for each agent, which is consistent with sublinear regret. This approach serves as a practical surrogate for tracking regret sublinearity, since computing the true regret is challenging without access to the time-varying optimal actions.}

\textcolor{blue}{Figures 2 and 3 show the effect of the learning rate at $\epsilon = 0.2$ and no communication or feedback delays. In Figure 3, the learning rate increases ten times. It can be seen that the average loss converges faster and closer to zero. This shows that an increase in the learning rate speeds up convergence.}

\textcolor{blue}{Figure 4 shows the effect of increasing privacy when $\epsilon = 0.1$. The learning rate remains unchanged, and there are no communication delays and feedback delays. It can be seen in Figure 4 that the average loss converges farther from zero than when $\epsilon = 0.2$ due to increased noise distortion.}   

\textcolor{blue}{Figure 5 shows the effect of a fixed communication delay when agent 2 communicates with agent 4, with no feedback delays. We impose the delay $\tau_{ij}(t)=\tau_{ij}=2$  and allow $\tau_i(t)=0 \quad\forall i\in\mathcal{V}$. The average loss of agent 4, which incurs the most loss among all agents, is lower in Figure 5 than in Figures 2 and 4 for the same learning rate. This is because the communication delay is fixed and occurs only on the odd iterations' communication link.}

\begin{figure}
     \centering
     \begin{subfigure}{0.22\textwidth}
         \centering
         \includegraphics[width=\linewidth]{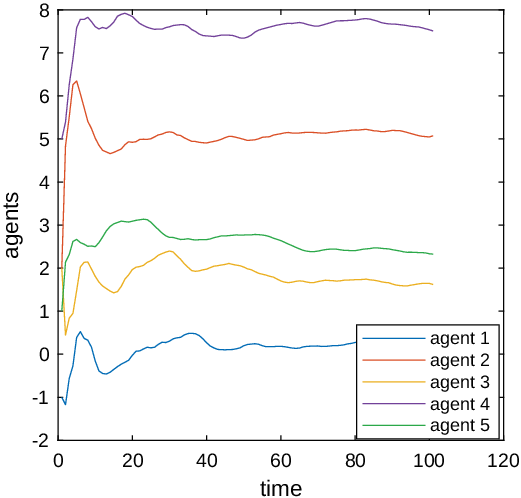}
         \label{(a)}
     \end{subfigure}
     \begin{subfigure}{0.23\textwidth}
         \centering
        \includegraphics[width=\linewidth, height = 1.5 in]{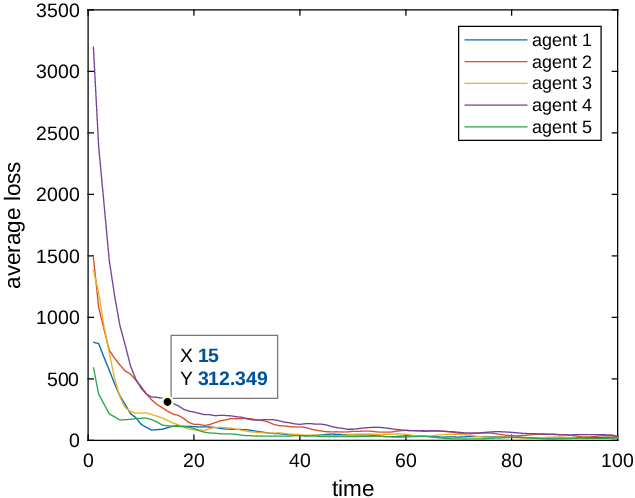}
         \label{(b)}
     \end{subfigure}
     \caption{\textcolor{blue}{ Trajectories of $x_i(t)$  (agents' actions) and $\frac{1}{t}\sum_{i=1}^t F_{i,t}(x_i(t),\psi(t))$ (average loss) with no communication and feedback delays, $\epsilon = 0.2$, $\eta_t = \frac{\gamma}{\sqrt{t+1}}$. 
     }}
     \label{fig:1}
\end{figure}

\begin{figure}
     \centering
     \begin{subfigure}{0.22\textwidth}
         \centering         \includegraphics[width=\linewidth]{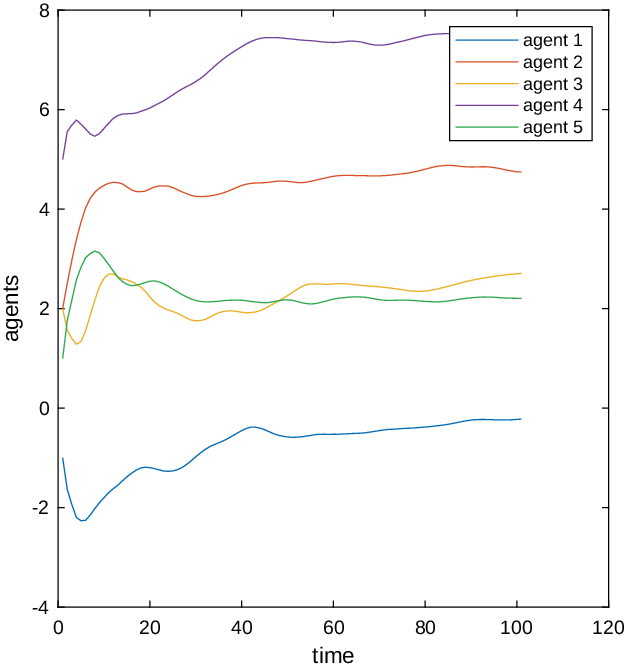}
         \label{(a)}
     \end{subfigure}
     \begin{subfigure}{0.23\textwidth}
         \centering       \includegraphics[width=\linewidth, height = 1.7 in]{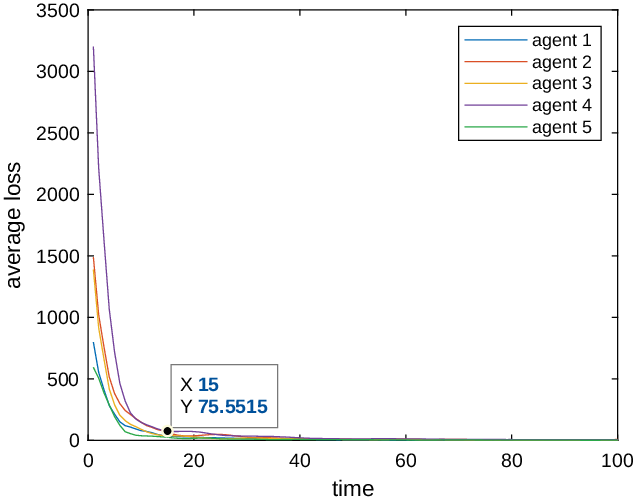}
         \label{(b)}
     \end{subfigure}
     \caption{\textcolor{blue}{(effect of increasing learning rate) Trajectories of $x_i(t)$  (agents' actions) and $\frac{1}{t}\sum_{i=1}^t F_{i,t}(x_i(t),\psi(t))$ (average loss) with no communication and feedback delays, $\epsilon = 0.2$, and $\eta_t = 10*\gamma/\sqrt{t+1}$. 
     }}
    \label{fig:1}
\end{figure}

\begin{figure}
     \centering
     \begin{subfigure}{0.22\textwidth}
         \centering
         \includegraphics[width=\linewidth]{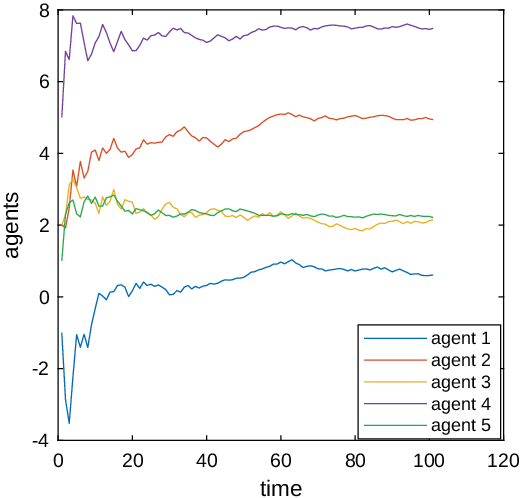}
         \label{(a)}
     \end{subfigure}
     \begin{subfigure}{0.23\textwidth}
         \centering
        \includegraphics[width=\linewidth, height = 1.5 in]{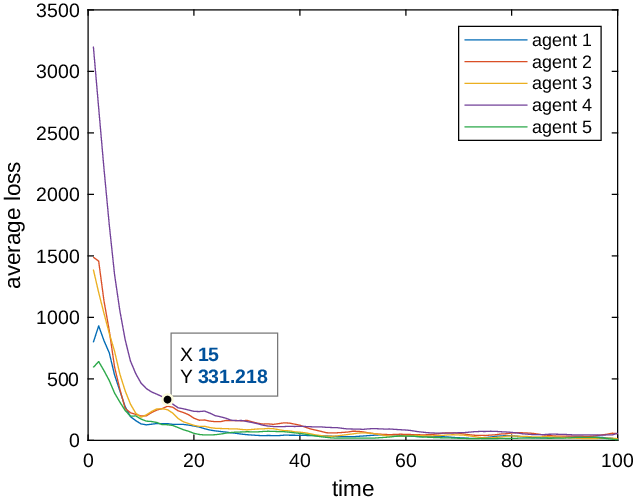}
         \label{(b)}
     \end{subfigure}
     \caption{\textcolor{blue}{(effect of increasing privacy) Trajectories of $x_i(t)$  (agents' actions) and $\frac{1}{t}\sum_{i=1}^t F_{i,t}(x_i(t),\psi(t))$ (average loss) with no communication and feedback delays, and $\epsilon = 0.1$, and $\eta_t = \gamma/\sqrt{t+1}$. 
     }}
     \label{fig:1}
\end{figure}

\begin{figure}
     \centering
     \begin{subfigure}{0.22\textwidth}
         \centering
         \includegraphics[width=\linewidth]{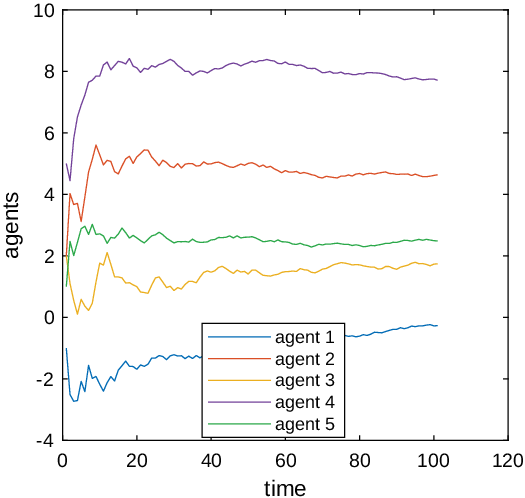}
         \label{(a)}
     \end{subfigure}
     \begin{subfigure}{0.23\textwidth}
         \centering
        \includegraphics[width=\linewidth, height=1.5 in]{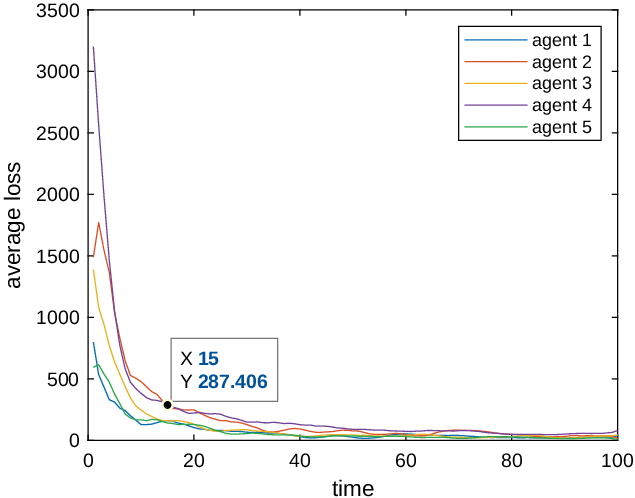}
         \label{(b)}
     \end{subfigure}
     \caption{\textcolor{blue}{(effect of fixed communication delays with no privacy) Trajectories of $x_i(t)$  (agents' actions) and $\frac{1}{t}\sum_{i=1}^t F_{i,t}(x_i(t),\psi(t))$ (average loss) with $\tau_i(t) = 0 \quad\forall i$ and $\tau_{42}(t)=\tau_{42} = 2
     $ and no privacy. 
     }}
     \label{fig:1}
\end{figure}


\begin{figure}
     \centering
     \begin{subfigure}{0.22\textwidth}
         \centering
         \includegraphics[width=\linewidth]{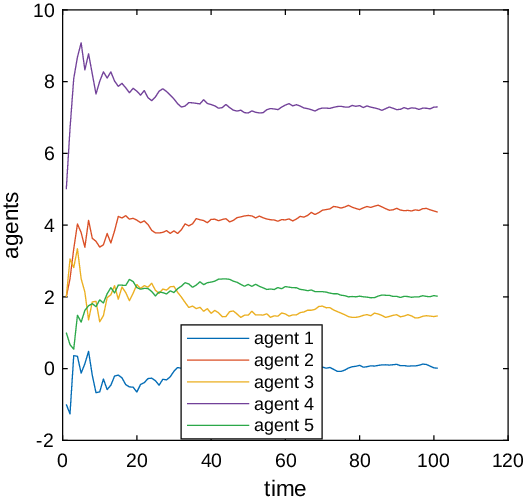}
         \label{(a)}
     \end{subfigure}
     \begin{subfigure}{0.23\textwidth}
         \centering
        \includegraphics[width=\linewidth, height = 1.5 in]{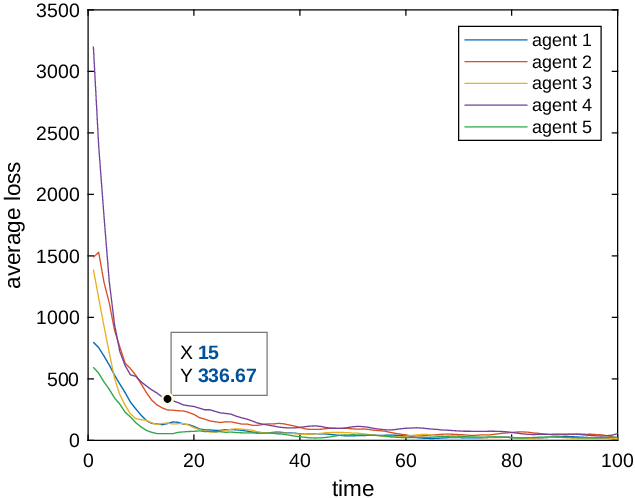}
         \label{(b)}
     \end{subfigure}
     \caption{\textcolor{blue}{(effect of time-varying and non-identical communication and feedback delays only) Trajectories of $x_i(t)$  (agents' actions) and $\frac{1}{t}\sum_{i=1}^t F_{i,t}(x_i(t),\psi(t))$ (average loss) with $\tau_i(t)\in (0,10)\quad\forall i\in\mathcal{V}$, and $\tau_{42}(t)\in(0,10)$ and no privacy. 
     }}
     \label{fig:1}
\end{figure}

\begin{figure}
     \centering
     \begin{subfigure}{0.22\textwidth}
         \centering
         \includegraphics[width=\linewidth]{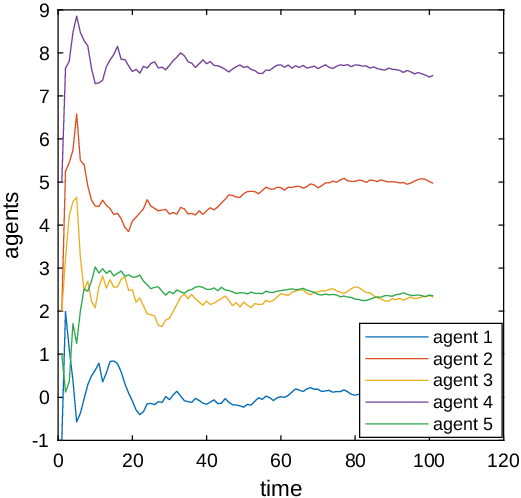}
         \label{(a)}
     \end{subfigure}
     \begin{subfigure}{0.23\textwidth}
         \centering
        \includegraphics[width=\linewidth, height = 1.5 in]{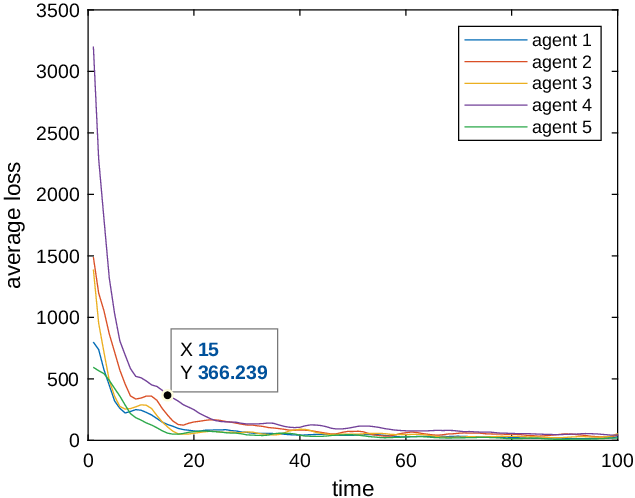}
         \label{(b)}
     \end{subfigure}
     \caption{\textcolor{blue}{(effect of time-varying and non-identical communication and feedback delays with privacy) Trajectories of $x_i(t)$  (agents' actions) and $\frac{1}{t}\sum_{i=1}^t F_{i,t}(x_i(t),\psi(t))$ (average loss) with $\tau_i(t)\in(0,10) \quad \forall i\in\mathcal{V}$,  $\tau_{42}(t)\in(0,10)$, and $\epsilon = 0.2$. 
     }}
     \label{fig:1}
\end{figure}

\begin{figure}
     \centering
     \begin{subfigure}{0.22\textwidth}
         \centering
         \includegraphics[width=\linewidth, height = 1.5 in]{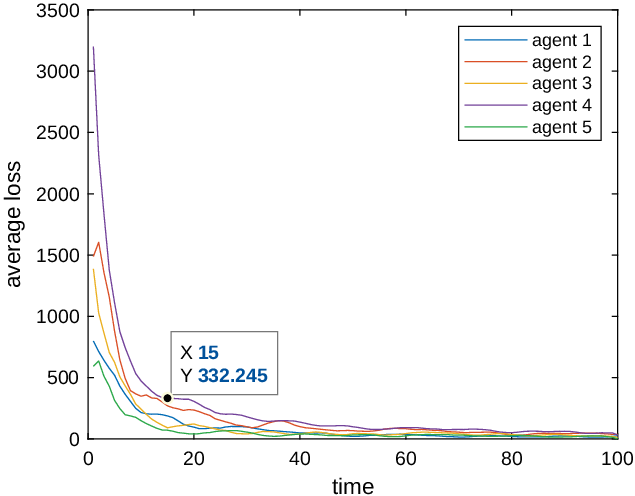}
         \label{(a)}
     \end{subfigure}
     \begin{subfigure}{0.23\textwidth}
         \centering
        \includegraphics[width=\linewidth, height = 1.5 in]{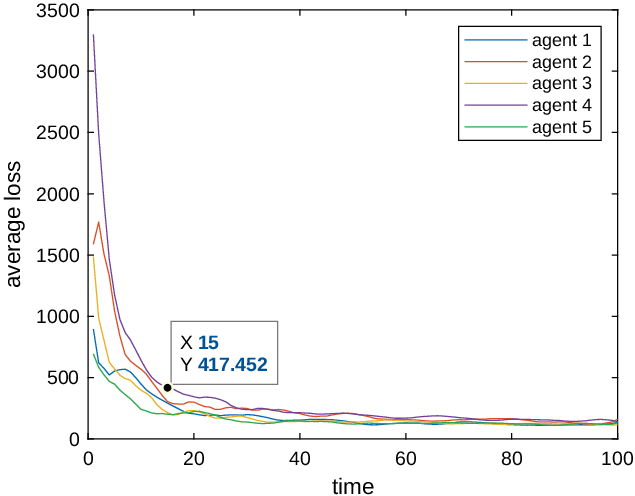}
         \label{(b)}
     \end{subfigure}
     \caption{\textcolor{blue}{(comparison of our proposed algorithm with \cite{liu2024sub}) Trajectories of $\frac{1}{t}\sum_{i=1}^t F_{i,t}(x_i(t),\psi(t))$ (average loss) with $\tau_i(t) =2 \quad\forall i\in\mathcal{V},\tau_{42} = 2$ and no privacy. 
    } }
     \label{fig:1}
\end{figure}

\textcolor{blue}{Figure 6 shows the effect of time-varying and non-identical communication and feedback delays but without privacy considerations.
We let $\tau_{ij}(t)$ and $\tau_i(t)$ for each firm be drawn from (0, 10) randomly at each time with $\tilde{\tau}=10$.  It can be seen in Figure 6 that the average loss converges for all agents, although the convergence is worse than in Figure 5. This is because agent 4 experiences a time-varying communication delay. Moreover, every agent, including agent 4, experiences time-varying and non-identical communication feedback delays.} 

\textcolor{blue}{Figure 7 shows the coupling effect of privacy, time-varying, and non-identical communication delays and feedback delays. By letting $\epsilon = 0.2$, it can be seen in Figure 7 that the average loss converges, but with a slightly worse performance compared to Figure 6. This is because of the coupling effect of privacy and delays. However, the convergence is expected to be better if we reduce the maximum noise variance, which is directly proportional to the sensitivity, as depicted in Theorems 1 and 2.}

\textcolor{blue}{Figure 8 compares the average loss incurred by the proposed Algorithm 1 and the state-of-the-art algorithm proposed in \cite{liu2024sub} for fixed communication and feedback delays and without privacy considerations. It can be seen that our proposed Algorithm 1 incurs lower average losses across the agents over time compared to the algorithm proposed in \cite{liu2024sub}.}


\section{Conclusion}
This paper proposed a privacy-preserving and delay-tolerant distributed dual averaging algorithm that achieves a time-varying Nash equilibrium in online learning settings. The proposed algorithm ensures privacy preservation of sensitive parameters of an agent during interaction with its time-varying in-neighbors in a time-varying unbalanced graph network. Moreover, the proposed algorithm effectively copes with time-varying and nonidentical communication delays and feedback delays, while still achieving a sublinear regret. The simulation result showed the convergence of the running average of each client's local action.

\appendices

\section{Proof of Lemma 3}
For simplicity, let $\mathbf{g}_{i,t-\tau_{i}(t)}(\mathbf{x}_i(t-\tau_{i}(t)),\mathbf{v}_i(t-\tau_{i}(t)))\equiv \mathbf{g}_{i,t-\tau_i(t)}$. From the definition of the average of the dual variable $\bar{\mathbf{b}}(t)=\sum_{i=1}^{V}\pi_i\mathbf{{b}}_i(t)$, we have
\begin{equation}
\begin{split}
   \bar{\mathbf{b}}(t+1) = \sum_{i=1}^{{V}} \pi_i(t+1)\bigg(\sum_{j=1}^{{V}}[{\mathbf{W}}^\prime(t)]_{ij}{\mathbf{{b}}}_j(t) + \frac{\mathbf{g}_{i,t-\tau_i(t)}}{y_{ii}(t)}\bigg) \\
=\sum_{j=1}^{V}{\mathbf{{b}}}_j(t) \bigg(\sum_{i=1}^{V}\pi_i(t+1)[\mathbf{W}^\prime(t)]_{ij}\bigg) +  \sum_{i=1}^{V}\pi_i(t+1)\\
\times\frac{\mathbf{g}_{i,t-\tau_i(t)}}{y_{ii}(t)}\\
= \sum_{j=1}^{V}{\mathbf{{b}}}_j(t)\pi_j(t) + \sum_{j=1}^{V} \pi_i(t+1)\frac{\mathbf{g}_{i,t-\tau_i(t)}}{y_{ii}(t)}\\
= \bar{\mathbf{b}}(t) + \sum_{i=1}^V \pi_i(t) \mathbf{n}_i(t) + \sum_{i=1}^{V}\pi_i(t+1)\frac{\mathbf{g}_{i,t-\tau_i(t)}}{y_{ii}(t)}.
\label{eqn 11}
   \end{split}
\end{equation}
The third equality is obtained from Lemma 1 and the row stochasticity of the augmented matrix $\mathbf{W}^\prime(t)$.

\section{Proof of Lemma 4}
Leveraging the definition of $\Pi^{\phi}_{\mathcal{X}}(\mathbf{b},\eta)=\arg\min_{\mathbf{x}\in\mathcal{X}}\{\langle \mathbf{b},\mathbf{x}\rangle + \frac{\phi(\mathbf{x})}{\eta}\bigg\} $
and $\mathbf{z}(t) := \Pi^{\phi}_{\mathcal{X}}(\bar{\mathbf{b}}(t),\eta_t)$, we have 
$\mathbf{z}(t-1)=\arg\min_{\mathbf{z}(t)\in \mathcal{X}}\bigg\{\langle \bar{\mathbf{b}}(t-1),\mathbf{z}(t)\rangle + \frac{1}{\eta_{t-1}}\phi(\mathbf{z}(t))\bigg\}$. Also, from Lemma 2, we have

\begin{equation}
\begin{split}
 \langle \bar{\mathbf{b}}(t-1),\mathbf{z}(t)\rangle + \frac{1}{\eta_{t-1}}\phi(\mathbf{z}(t))\geq \\ \langle \bar{\mathbf{b}}(t-1),\mathbf{z}(t-1)\rangle +\frac{1}{\eta_{t-1}}\phi(\mathbf{z}(t-1))+\\ \frac{1}{\eta_{t-1}}B_{\phi}(\mathbf{z}(t),\mathbf{z}(t-1)), 
 \end{split}
\end{equation}
thus, we have
\begin{equation}
\begin{split}
 -\langle \bar{\mathbf{b}}(t-1),\mathbf{z}(t)\rangle \leq -\langle \bar{\mathbf{b}}(t-1),\mathbf{z}(t-1)\rangle +\\ \frac{1}{\eta_{t-1}}[\phi(\mathbf{z}(t))-\phi(\mathbf{z}(t-1))] - \frac{1}{\eta_{t-1}}B_{\phi}(\mathbf{z}(t),\mathbf{z}(t-1)).  
 \end{split}
 \label{eqn 17*}
\end{equation}
Leveraging Lemma 3, we have $\bar{\mathbf{b}}(t)= \bar{\mathbf{b}}(t-1) + \sum_{i=1}^{V} \pi_i(t)\frac{\mathbf{g}_{i,t-\tau_i(t)-1}}{y_{ii}(t-1)}$, and $B_{\phi}(\mathbf{z}(t),\mathbf{z}(t-1))\geq 0$, we obtain 

\begin{equation}
\begin{split}
 \sum_{i=1}^{V}\pi_i(t)\langle \frac{\mathbf{g}_{i,t-\tau_i(t)-1}}{y_{ii}(t-1)}, \mathbf{z}(t)-\mathbf{x}_i^*(t)\rangle\\
 = \langle \bar{\mathbf{b}}(t),\mathbf{z}(t)-\mathbf{x}^*_i(t)\rangle - \langle \bar{\mathbf{b}}(t-1),\mathbf{z}(t)-\mathbf{x}^*_i(t)\rangle\\
 \leq \langle \bar{\mathbf{b}}(t),\mathbf{z}(t)-\mathbf{x}^*_i(t)\rangle - \langle \bar{\mathbf{b}}(t-1),\mathbf{z}(t-1)-\mathbf{x}^*_i(t-1)\rangle\\
 + \frac{1}{\eta_{t-1}}[\phi(\mathbf{z}(t))-\phi(\mathbf{z}(t-1))],
 \end{split}
\end{equation}
where the inequality is obtained from (\ref{eqn 17*}). Hence,
\begin{equation*}
\begin{split}
\sum_{t=1}^T \sum_{i=1}^{V} \pi_i(t) \langle \frac{\mathbf{g}_{i,t-\tau_i(t)-1}}{y_{ii}(t-1)}, \mathbf{z}(t)-\mathbf{x}^*_i(t)\rangle\\
\leq \langle \bar{\mathbf{b}}(T),\mathbf{z}(T)-\mathbf{x}^*_i(T)\rangle + \sum_{t=2}^T \bigg[\frac{1}{\eta_{t-2}}-\frac{1}{\eta_{t-1}}\bigg]\phi(\mathbf{z}(t-1))\\
+ \frac{1}{\eta_{T-1}}\phi(\mathbf{z}(T))
\end{split}
\end{equation*}
\begin{equation}
 \begin{split}
 =\langle \bar{\mathbf{b}}(T),\mathbf{z}(T)\rangle + \frac{1}{\eta_T}\phi(\mathbf{z}(T))-\langle \bar{\mathbf{b}}(T),\mathbf{x}^*_i(T)\rangle -\\ \frac{1}{\eta_T}\phi(\mathbf{x}^*_i(T))+\frac{1}{\eta_T}\phi(\mathbf{x}^*_i(T))+\bigg[\frac{1}{\eta_{T-1}}-\frac{1}{\eta_T}\bigg]\phi(\mathbf{z}(T))\\
 + \sum_{t=2}^T\bigg[\frac{1}{\eta_{t-2}}-\frac{1}{\eta_{t-1}}\bigg]\phi(\mathbf{z}(t-1)).
\end{split}
\end{equation}
Because $\mathbf{z}(T)=\arg\min_{\mathbf{z}\in\mathbf{X}}\{\langle \bar{\mathbf{b}}(T),\mathbf{z}\rangle + \frac{1}{\eta_T}\phi(\mathbf{z})\}$, it follows that
\begin{equation}
\begin{split}
\sum_{t=1}^T \sum_{i=1}^{V}\pi_i(t) \langle \frac{g_{i,t-\tau_i(t)-1}}{y_{ii}(t-1)},\mathbf{z}(t)-\mathbf{x}^*_i(t)\rangle \\
\leq \frac{\phi(\mathbf{x}^*_i(T))}{\eta_T} + \bigg[\frac{1}{\eta_{T-1}}-\frac{1}{\eta_T}\bigg]\phi(\mathbf{z}(T)) \\+ \sum_{t=2}^T \bigg[\frac{1}{\eta_{t-2}}-\frac{1}{\eta_{t-1}}\bigg]\phi(\mathbf{z}(t-1)).  
\end{split}
\end{equation}
Because $\eta_t$ is nonincreasing and $\phi(\cdot)\geq 0$, we have
\begin{equation}
 \sum_{t=1}^T \sum_{i=1}^{V}\pi_i(t)\langle\frac{\mathbf{g}_{i,t-\tau_i(t)-1}}{y_{ii}(t-1)},\mathbf{z}(t)-\mathbf{x}^*_i(t)\rangle \leq \frac{\phi(\mathbf{x}^*_i(T))}{\eta_T}.    
\end{equation}

\section{Proof of Theorem 1}
Using the recursion of Step 7 in Algorithm 1, we have

\begin{equation}
 \mathbf{{b}}_i(t) = \sum_{j=1}^{V} \sum_{s=1}^{t-1} \bigg[\mathbf{W}^\prime(t-1:s+1)\bigg]_{ij} \bigg (\mathbf{n}_j(t) + \frac{\mathbf{g}_{j,s-\tau_j(t)}}{y_{jj}(t)}\bigg)    
 \label{eqn 12}
\end{equation}

Using (\ref{eqn 12}), we have
\begin{equation*}
    \bar{\mathbf{b}}_t = \sum_{i=1}^{V}\pi_i(t)\mathbf{{b}}_i(t) 
\end{equation*}
\begin{equation*}
\begin{split}
  =\sum_{i=1}^{V} \pi_i(t) \sum_{j=1}^{V}\sum_{s=1}^{t-1}\bigg[\mathbf{W}^\prime(t-1:s+1)\bigg]_{ij}\times \\
  \bigg(\mathbf{n}_j(t) +\frac{\mathbf{g}_{j,s-\tau_j(s)}}{y_{jj}(s)}\bigg)
  \end{split}
  \end{equation*}
  \begin{equation*}
  \begin{split}
 = \sum_{j=1}^{V}\sum_{s=1}^{t-1}\bigg(\mathbf{n}_j(t) +\frac{\mathbf{g}_{j,s-\tau_j(t)}}{y_{jj}(s)}\bigg)\times\\ \sum_{i=1}^{V} \pi_i(t) \bigg[\mathbf{W}^\prime(t-1:s+1)\bigg]_{ij}
 \end{split}
 \end{equation*}
 \begin{equation}
 \begin{split}
 =\sum_{j=1}^{V}\sum_{s=1}^{t-1} \pi_j(s+1)\bigg(\mathbf{n}_j(t) +\frac{\mathbf{g}_{j,s-\tau_j(t)}}{y_{jj}(s)}\bigg),
 \end{split}
 \label{eqn 13}
\end{equation}
and 
\begin{equation}
 \mathbf{z}(t) = \Pi_{\mathcal{X}}^{\phi}\bigg(\sum_{j=1}^{V}\sum_{s=1}^{t-1}\pi_j(s+1)\frac{\mathbf{g}_{j,s-\tau_j(s)}}{y_{jj}(s)}+\mathbf{n}_j(t),\eta_t\bigg).   
\end{equation}
Let $F_{i,t}(\mathbf{x}_i(t),\Psi(\mathbf{x}_i(t),\mathbf{x}^*_{-i}(t)))\equiv F_{i,t}(\mathbf{x}_i(t)))$ and $F_{i,t}(\mathbf{x}^*_i(t),\Psi(\mathbf{x}^*(t))\equiv F_{i,t}(\mathbf{x}_i^*(t))$ for simplicity.
\begin{equation}
\begin{split}
\sum_{t=1}^T\sum_{i=1}^{V} F_{i,t}({\mathbf{x}}_i(t)) - \sum_{t=1}^T\sum_{i=1}^{V} F_{i,t}(\mathbf{x}^*_i(t)) \\ = 
\sum_{t=1}^T \sum_{i=1}^{V} \bigg[F_{i,t}(\mathbf{z}(t))-F_{i,t}(\mathbf{x}^*_i(t))\bigg] +  \\ \sum_{t=1}^T\sum_{i=1}^{V} \bigg[F_{i,t}(\mathbf{x}_i(t))-F_{i,t}(\mathbf{z}(t))\bigg] 
\\ \leq 
\sum_{t=1}^T \sum_{i=1}^{V} \bigg[F_{i,t}(\mathbf{z}(t))-F_{i,t}(\mathbf{x}^*_i(t))\bigg] +\\ L\sum_{t=1}^T \sum_{i=1}^{V} \eta_t ||\mathbf{b}_i(t) - \bar{\mathbf{b}}(t)||_*. 
\end{split}
\label{eqn 15}
\end{equation}
We leverage both Assumption 4 and Lemma 5 in the inequality in (\ref{eqn 15}). Bounding the first term of the last inequality in (\ref{eqn 15}) gives
\begin{equation}
\begin{split}
   \sum_{t=1}^T \sum_{i=1}^{V} \bigg[F_{i,t}(\mathbf{z}(t))-F_{i,t}(\mathbf{x}^*_i(t))\bigg]\\=
   \sum_{t=1}^T \sum_{i=1}^{V}[F_{i,t}(\mathbf{z}(t))- F_{i,t}(\mathbf{x}_i(t))]+\\ \sum_{t=1}^T \sum_{j=1}^{V} [F_{i,t}(\mathbf{x}_i(t))-F_{i,t}(\mathbf{x}^*_i(t))]
   \\ \leq L\sum_{t=1}^T \sum_{j=1}^{V}\eta_t ||\bar{\mathbf{b}}(t)-\mathbf{b}_i(t)||_* +\\
   \sum_{t=1}^T \sum_{i=1}^{V}\langle \mathbf{g}_{i,t}, \mathbf{x}_i(t)-\mathbf{x}^*_i(t)\rangle,
   \label{eqn 16}
   \end{split}
\end{equation}
the last inequality is obtained from Assumption 4 and Lemma 5, and convexity. Let us define the error in the gradient as $\mathbf{e}_i(t) :=\mathbf{g}_{i,t}-\mathbf{g}_{i,t-\tau_i(t)-1}$, then 
\begin{equation}
\begin{split}
   \langle \mathbf{g}_{i,t},\mathbf{x}_i(t)-\mathbf{x}^*_i(t)\rangle  \\
   = \langle \mathbf{g}_{i,t-\tau_i(t)-1},\mathbf{z}(t)-\mathbf{x}^*_i(t)\rangle + \langle \mathbf{e}_i(t), \mathbf{x}_i(t)-\mathbf{x}^*_i(t)\rangle +\\ \langle \mathbf{g}_{i,t-\tau_i(t)-1},\mathbf{x}_i(t)-\mathbf{z}(t)\rangle\\
   \leq \langle \mathbf{g}_{i,t-\tau_i(t)-1}, \mathbf{z}(t)-\mathbf{x}^*_i(t)\rangle + \langle \mathbf{e}_i(t), \mathbf{x}_i(t)-\mathbf{x}^*_i(t)\rangle +\\
   L\eta_t ||\mathbf{b}_i(t)-\bar{\mathbf{b}}_t||_*,
   \end{split}
   \label{eqn 17}
\end{equation}
We leverage Assumption 4 and Lemma 5 in (\ref{eqn 17}).
Using the three point identity of Bregman's divergence  $\langle \triangledown F(\mathbf{a})-\triangledown F(\mathbf{b}), \mathbf{b}-\mathbf{c}\rangle = B_F(\mathbf{c},\mathbf{a})-B_F(\mathbf{c},\mathbf{b})-B_F(\mathbf{b},\mathbf{a})$, we can rewrite $\langle \mathbf{e}_{i}(t), \mathbf{x}_i(t)-\mathbf{x}^*_i(t)\rangle$ as
\begin{equation}
\begin{split}
    \langle \mathbf{e}_i(t), \mathbf{x}_i(t)-\mathbf{x}^*_i(t)\rangle \\
    = -\langle \mathbf{g}_{i,t-\tau_i(t)-1}-\mathbf{g}_i(t),\mathbf{x}_i(t)-\mathbf{x}^*_i(t)\rangle\\
    =-B_{F_{i,t}}\langle \mathbf{x}^*_i(t), \mathbf{x}_i(t-\tau_i(t)-1) + B_{F_{i,t}}(\mathbf{x}^*_i(t),\mathbf{x}_i(t)\rangle\\
    + B_{F_{i,t}}(\mathbf{x}_i(t),\mathbf{x}_i(t-\tau_i(t)-1)).
    \label{eqn 18}
    \end{split}
\end{equation}
Summing over the time horizon gives
\begin{equation*}
  \sum_{t=1}^T \langle \mathbf{e}_i(t), \mathbf{x}_i(t)-\mathbf{x}^*_i(t)\rangle
  \end{equation*}
  \begin{equation*}
  \begin{split}
  \leq \sum_{t=T-\bar{\tau}}^T B_{F_{i,t}}(\mathbf{x}^*_i(t), \mathbf{x}_i(t)) +
  \sum_{t=1}^T B_{F_{i,t}}(\mathbf{x}_i(t),\mathbf{x}_i(t-\tau_i(t)-1))
  \end{split}
\end{equation*}
\begin{equation}
\begin{split}
 \leq \sum_{t=T-\bar{\tau}}^T 2L||\mathbf{x}_i(t)-\mathbf{x}^*_i(t)||+
 \sum_{t=1}^T \frac{G}{2}||\mathbf{x}_i(t)-\mathbf{x}_i(t-\\\tau_i(t)-1)||^2\\
 \leq 2(\bar{\tau}+1)LR + \sum_{t=1}^T \frac{G}{2}||\mathbf{x}_i(t)-\mathbf{x}_i(t-\tau_i(t)-1)||^2,
 \end{split}
 \label{eqn 19}
\end{equation}
We obtain the second inequality from bounding the definition of $B_{F_{i,t}}(\mathbf{x}^*(t), \mathbf{x}_i(t))$ with Assumption 4(a) and also applying Assumption 4(b), and the last inequality from Assumption 5. We can further bound the second term of the last inequality in (\ref{eqn 19}) using $(A+B+C)^2\leq 3(A^2 + B^2 + C^2)$; thus, we have
\begin{equation*}
 ||\mathbf{x}_i(t)-\mathbf{x}_i(t-\tau_i(t)-1)||^2 
 \end{equation*}
 \begin{equation*}
 \begin{split}
 \leq 3(||\mathbf{x}_i(t)-\mathbf{z}(t)||^2 + ||\mathbf{z}(t)-\mathbf{z}(t-\tau_i(t)-1)^2||\\
 +||\mathbf{x}_i(t-\tau_i(t)-1) - \mathbf{z}(t-\tau_i(t)-1||^2)
 \end{split}
 \end{equation*}
 \begin{equation*}
 \begin{split}
 \leq 3\eta_t^2||\mathbf{b}_i(t) - \bar{\mathbf{b}}(t)||^2_* + 3||\mathbf{z}(t)-\mathbf{z}(t-\tau_i(t)-1)||^2\\
 + 3\eta_t^2(t-\tau_i(t)-1)||\mathbf{b}_i(t-\tau_i(t)-1)-\bar{\mathbf{b}}(t)||_*^2
 \end{split}
 \end{equation*}
 \begin{equation}
 \begin{split}
 \leq 3\eta_t^2||\mathbf{b}_i(t)-\bar{\mathbf{b}}(t)||_*^2 + 3\eta_t^2(t-\tau_i(t)-1)(\tilde{\tau}+1)^2L^2\theta^2\\
 +3\eta_t^2(t-\tau_i(t)-1) ||\mathbf{b}_i(t)-\bar{\mathbf{b}}(t)||^2_*
 \end{split}
 \label{eqn 20}
\end{equation}
we leverage Lemmas 1(a),  3 and 5 in the last inequality in (\ref{eqn 20}). Furthermore, bounding the first term of (\ref{eqn 17}) and summing over time and agents, we have
\begin{equation*}
\begin{split}
  \sum_{i=1}^T\sum_{i=1}^{V}\bigg\langle \mathbf{g}_{i,t-\tau_i(t)-1}, \mathbf{z}(t)-\mathbf{x}^*_i(t)\bigg\rangle  \\
  = \sum_{t=1}^T \sum_{i=1}^{V}
  \bigg\langle \frac{\pi_i(t)\cdot \mathbf{g}_{i,t-\tau_i(t)-1}\cdot y_{ii}(t-1)}{\pi_i(t)\cdot y_{ii}(t-1)},\\ \mathbf{z}(t)-\mathbf{x}^*_i(t)\bigg\rangle\\
  \leq \frac{1}{\beta} \sum_{t=1}^T \sum_{i=1}^{V} \bigg\langle \frac{\pi_i(t)\cdot \mathbf{g}_{i,t-\tau_i(t)-1}}{y_{ii}(t-1)},\\
  \mathbf{z}(t)-\mathbf{x}^*_i(t))\bigg\rangle
  \end{split}
\end{equation*}
\begin{equation}
    \leq \frac{\phi(\mathbf{x}^*_i(T))}{\beta\eta_T}.
    \label{eqn 21}
\end{equation}
The first inequality is based on Lemma 1, and the second inequality is based on Lemma 4. We can subtract (\ref{eqn 12}) from (\ref{eqn 13}) to bound the first and third term in (\ref{eqn 20}) to  obtain
\begin{equation*}
\begin{split}
\bar{\mathbf{b}}(t)-\mathbf{b}_i(t) = \sum_{s=1}^{t-1}\sum_{j=1}^{V} \bigg(\pi_j(s+1)-  [W^\prime(t-1:s+1)]_{ij}\bigg)\\ \times
\bigg(\frac{\mathbf{g}_{j,s-\tau_j(s)}}{y_{jj}(s)}+\mathbf{n}_j(t)\bigg)
\end{split}
\end{equation*}
Taking the dual norm gives
\begin{equation*}
\begin{split}
||\bar{\mathbf{b}}(t)-\mathbf{b}_i(t)||_* \leq \sum_{s=1}^{t-1}\sum_{j=1}^{V} \bigg|\pi_j(s+1)-  [W^\prime(t-1:s+1)]_{ij}\bigg|\\ 
\times
\bigg|\bigg|\frac{\mathbf{g}_{j,s-\tau_j(s)}}{y_{jj}(s)}\bigg|\bigg|_* \times \mathbb{E}[||\mathbf{n}_j(t)||]
\end{split}
\end{equation*}
\begin{equation*}
\leq \sqrt{2m}C{V}\sigma_{t} L\theta \sum_{s=1}^{t-1}{\lambda}^{t-s-2}
\end{equation*}
\begin{equation}
\leq \frac{\sqrt{2m} CV L \theta \sigma_{max}}{\lambda(1-\lambda)}, 
\label{eqn 22}
\end{equation}
The second inequality follows from Lemma 1 and $\mathbb{E}[||\mathbf{n}(t)||^2]=2m\sigma^2(t)$. The third inequality leverages the sum to infinity and $\max_{0\leq t\leq T}\sigma(t) = \sigma_{max}$
By combining (\ref{eqn 15}), (\ref{eqn 16})-(\ref{eqn 17}), (\ref{eqn 19})-(\ref{eqn 20}), (\ref{eqn 21}) and (\ref{eqn 22}), we have
\begin{equation}
\begin{split}
 \sum_{t=1}^T \sum_{j=1}^{V} F_{i,t}({\mathbf{x}}_{i}(t))  - \sum_{t=1}^T\sum_{j=1}^{V} F_{i,t}(\mathbf{x}^*_i(t))\\ \leq K_1 \frac{1}{\sqrt{T}} + K_2\frac{\ln T}{T} + K_3\frac{1}{T}.\\
  = O\bigg(\frac{\tilde{\tau}^2 V \sigma_{max}}{(1-\lambda)\sqrt{T}}\bigg),
  \end{split}
\end{equation}
where 
$\Delta_t = 2L\theta\sqrt{m}$ is the sensitivity,

$K_1 = \frac{4(\bar{\tau}+1)^2CVL^2\theta \gamma \sigma_{max}}{\lambda(1-\lambda)} + \frac{2(\bar{\tau}+1)CVL^2\theta\gamma}{\lambda(1-\lambda)}+\frac{R^2}{2VB\gamma}$,

$K_2 = \frac{3GV^3C^2L^2\theta\gamma}{V\lambda^2(1-\lambda)^2}+ \frac{3GV(\bar{\tau}+1)^2L^2\theta^2\gamma^2}{2V}$,

$K_3 = \frac{2GV(\bar{\tau}+1)LR}{V}+\frac{3GV^3C^2L^2\theta^2\gamma^2}{V\lambda^2(1-\lambda)^2} + \frac{3GV(\bar{\tau}+1)^2L^2\theta^2\gamma^2}{2N}$.

\section{Proof of Lemma 6}
From Definition 2, the two adjacent cost functions are $Z_{i,t}$ and $Z^\prime_{i,t}$. Let $Z_t:= \cup_{i=1,...,V}Z_{i,t}$ and $Z^\prime_t := \cup_{i=1,...,V}Z^\prime_{i,t}$. Under differential privacy, the adversary cannot differentiate $\tilde{\mathbf{b}}_{i,t}$ from $\tilde{\mathbf{b}}^\prime_{i,t}$. By leveraging Step 7 of Algorithm 1, we obtain
\begin{equation}
\begin{split}
    ||\mathbf{b}_{i,t+1} - \mathbf{b}^\prime_{i,t+1}||_1 \leq ||\frac{\mathbf{g}_{i,t-\tau_i(t)}}{y_{ii}(t)}||_1 + ||\frac{\mathbf{g}^\prime_{i,t-\tau_i(t)}}{y^\prime_{ii}(t)}||_1\\
    \leq \theta\sqrt{m}(||\mathbf{g}_{t-\tau_i(t)}|| + ||\mathbf{g}^\prime_{t-\tau_i(t)}||)\\
    \leq 2L\theta\sqrt{m},
    \end{split}
\end{equation}
The property of $1-$norm is used in the first inequality. From Definition 4, we have
\begin{equation}
\begin{split}
 \Delta_t =\sup_{{Z_{i,t},Z^\prime_{i,t}}:\text{Adj}(Z_{i,t},Z^\prime_{i,t})}||\mathcal{A}(Z_{i,t})-\mathcal{A}(Z^\prime_{i,t})||_1\\ \leq 2L\theta\sqrt{m}.
 \end{split}
\end{equation}
The sensitivity is dependent on the estimate $y_{ii}(t)$, the constant $L$, and the dimension of $\mathbf{b}_{i,t}\in\mathbb{R}^m$. 
$\indent\indent\indent\indent\indent\indent\indent\indent\indent\indent\indent\indent\indent\indent\indent\indent\indent\indent\indent\indent\indent\indent\indent\indent\blacksquare$
\section{Proof of Theorem 2}
Let us define $\mathbf{b}_t=[\mathbf{b}_{1,1},...\mathbf{b}_{1,t},...,\mathbf{b}_{V,t}]\in\mathbb{R}^{Vm}$ and $\mathbf{b}^\prime_t = [\mathbf{b}^\prime_{1,t},...,\mathbf{b}^\prime_{1,t},...,\mathbf{b}^\prime_{V,t}]\in\mathbb{R}^{Vm}$. Leveraging Definition 4, we have 
\begin{equation}
    ||\mathbf{b}_t-\mathbf{b}^\prime_t||_1 = \sum_{i=1}^V\sum_{k=1}^m |{b}^k_{i,t}-{b}^{\prime k}_{i,t}|\leq \Delta_t
\end{equation}
where $b^k_{i,t}$ and $b^{\prime k}_{i,t}$ are the $k-th$ components of $\mathbf{b}_{i,t}$ and $\mathbf{b}^{\prime}_{i,t}$ respectively. Since the adversary observes $\tilde{\mathbf{b}}_{i,t}$ and $\tilde{\mathbf{b}^\prime}_{i,t}$ as the same after randomizing with Laplace noise, then it is sufficient to show following Proposition 1 of \cite{xiong2020privacy} that
\begin{equation}
\begin{split}
    \prod_{i=1}^V \prod_{k=1}^m \frac{[\tilde{b}^k_{i,t}-b^k_{i,t}]}{[\tilde{b}^{\prime k}_{i,t}-b^k_{i,t}]}\\ \leq \exp\bigg(\frac{||\mathbf{b}^\prime_{i,t}-\mathbf{b}_{i,t}||_1}{\sigma_t}\bigg)\\
   \leq \exp\bigg(\frac{\Delta_t}{\sigma_t}\bigg) = e^{\epsilon_t}.
    \end{split}
\end{equation}
On the other hand, 
\begin{equation}
 \mathbb{P}[\mathcal{A}(Z)\in S] = \prod_{t=1}^T \mathbb{P}[\mathcal{A}(Z_t)\in S],  
\end{equation}
Thus, we can infer that
\begin{equation}
   \prod_{t=1}^T \mathbb{P}[\mathcal{A}(Z_t)\in S]\leq \prod_{t=1}^T \exp \bigg(\frac{\Delta_t}{\sigma_t}\bigg)\cdot  [\mathcal{A}(Z^\prime_t)\in S],
\end{equation}
which based on $\hat{\epsilon}=\sum_{t=1}^T (\Delta_t/\sigma_t)$ gives
\begin{equation}
  \mathbb{P}[\mathcal{A}(Z)\in S]\leq \exp(\hat{\epsilon})\cdot\mathbb{P}[\mathcal{A}(Z^\prime)\in S].  
\end{equation}


%





\ifCLASSOPTIONcaptionsoff
  \newpage
\fi



\bibliographystyle{IEEEtran}
\bibliography{Ref.bib}








\end{document}